\documentclass[aps,prapplied,twocolumn,superscriptaddress,floatfix]{revtex4-2}
\usepackage[T1]{fontenc}
\usepackage{graphicx}
\usepackage{amssymb,amsfonts,amsmath}
\usepackage[colorlinks=true,citecolor=Cerulean,linkcolor=RubineRed,urlcolor=Cerulean]{hyperref}
\hypersetup{breaklinks=true}
\usepackage{graphicx}
\usepackage{color}
\usepackage[usenames,dvipsnames]{xcolor}
\usepackage{epstopdf}
\usepackage[normalem]{ulem}
\usepackage{bm}
\usepackage{booktabs}
\usepackage{bbold}
\usepackage{float}
\usepackage{dsfont}
\usepackage{comment}
\usepackage{algorithm}
\usepackage{algpseudocode}
\usepackage{makecell}
\usepackage{multirow}

\newcommand{\ket}[1]{|#1\rangle}
\newcommand{\bra}[1]{\langle #1|}

\usepackage{mathtools}

\begin{document}

\title{Hybrid Quantum Repeater Chains with Semiconductor Quantum Dots and Group-IV-Vacancy Color Centers in Diamond}

\author{Yannick Strocka}
\altaffiliation{These authors contributed equally to this work}
\affiliation{Department of Physics, Humboldt-Universität zu Berlin, 12489 Berlin, Germany}%

\author{Fenglei Gu}
\altaffiliation{These authors contributed equally to this work}
\affiliation{QuTech, Technische Universiteit Delft, 2628 CD Delft, Netherlands}%

\author{Gregor Pieplow}
\affiliation{Department of Physics, Humboldt-Universität zu Berlin, 12489 Berlin, Germany}%

\author{Johannes Borregaard}
\affiliation{Department of Physics, Harvard University, Cambridge, Massachusetts 021388, USA}%

\author{Tim Schröder}
\affiliation{Department of Physics, Humboldt-Universität zu Berlin, 12489 Berlin, Germany}%
\affiliation{Ferdinand-Braun-Institut, Leibniz-Institut für Höchstfrequenztechnik, 12489 Berlin, Germany}%

\begin{abstract}
We propose and analyze a hybrid quantum repeater architecture that combines two leading hardware platforms: quantum dots (QDs) as bright, deterministic sources of entangled photon pairs and group-IV-vacancy centers in diamond as efficient, heralded quantum memories. This combination leverages high-rate entanglement generation together with long-lived storage, enabling scalable entanglement distribution over long distances. A key challenge is the large bandwidth mismatch between QD photons and the narrow optical transitions of the memories. We combine a comprehensive model of the spin-photon interface, including full spin-photon coupling dynamics,
and explore mitigation strategies such as frequency filtering and optimized magnetic-field orientation. Our results show that with optimized designs, photon-to-memory transfer can be achieved with high efficiency and fidelity, supporting the feasibility of such hybrid systems. Finally, we analyze a full repeater chain using experimentally achievable parameters and find that a network with thousands of memories across several repeater nodes could achieve a secret-key rate of $500$ bit/s over $1\textnormal{,}000$ km, demonstrating the strong potential of this approach for next-generation quantum networks.
\end{abstract}

\maketitle

\begin{figure*}[tb]
    \centering
    \includegraphics[width=\textwidth]{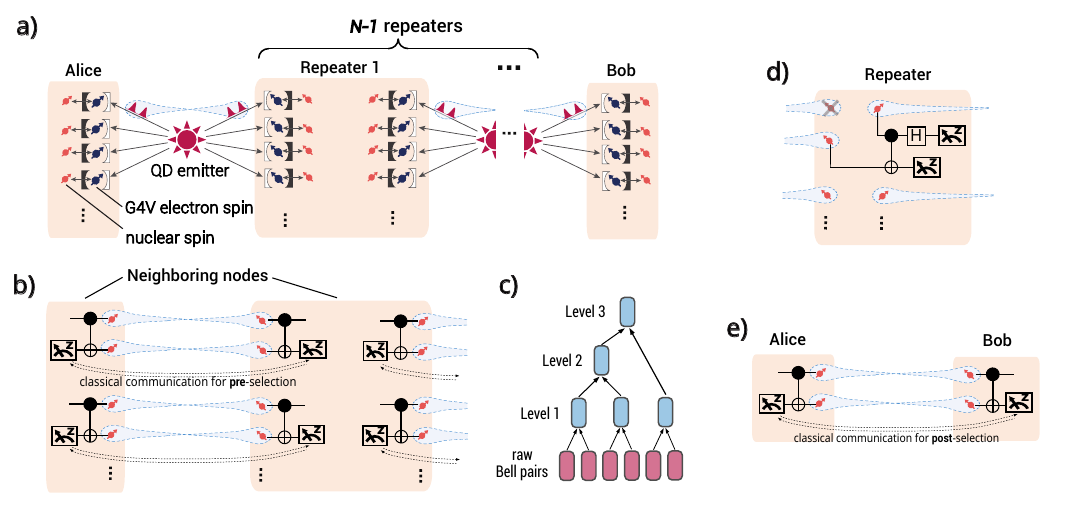}
    \caption{
    QD-G4V hybrid repeater chain protocol.
    \textbf{a)} Elementary-link establishment. The chain consists of two end nodes, Alice and Bob, and $N-1$ intermediate repeater nodes, each represented by an orange rectangle color block. QD emitters (red stars) are positioned midway between neighboring nodes and distribute entangled photons (red bell shapes) between them. In each node, G4V electron spins (dark blue) serve as interfaces for loading qubits from the incoming entangled photons, while nuclear spins (orange) provide long-term qubit storage.
    \textbf{b)} The fundamental operations of the elementary-link entanglement distillation. They involve local nucleus-nucleus \textsc{cnot} gates and qubit measurement and classical communications between the two nodes.
    \textbf{c)} The entanglement distillation protocol following Ref.~\cite{distillation_nigmatullin2016minimally}. It has four options: level-0: using directly the raw Bell pairs; level-1, -2, and -3: each distilled Bell pair is made up from 2, 4, and 6 raw ones, respectively. Single-qubit gates (not shown) are applied before the level-2 and -3 distillations.
    \textbf{d)} Entanglement swapping. It contains a \textsc{cnot} gate between the two nuclear spins from different memory modules in a repeater node, followed by a Hadamard gate and qubit measurements.
    \textbf{e)} The end-to-end entanglement distillation is similar to the elementary-link distillation, except that, due to its time-consuming nature, classical communication is employed for post-selection rather than as a precondition in the level-2 and level-3 operations.
    }
    \label{fig:overview1}
\end{figure*}

\begin{figure}[h!]
    \centering
    \includegraphics[width=\linewidth]{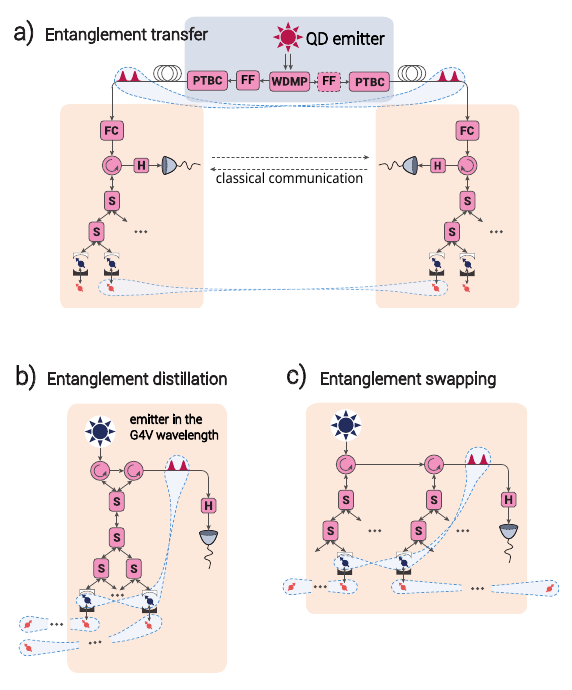}
    \caption{Optical circuits for the repeater-chain operations.
    \textbf{a)} Entanglement transfer from the photons to the nuclear spins as in the elementary-link establishment step. QDs emit frequency-distinct, polarization-entangled photon pairs, which are separated by a wavelength-division multiplexer (WDMP). Frequency filters (FFs) are used on one side or both for narrowing the bandwidth of the photon(s) in a Bell pair. Polarization-to-time-bin converters (PTBCs), implemented with polarizing beam splitters, encode polarization qubits into time-bin qubits for fiber transmission. At each repeater node, frequency converters (FCs) shift photons from telecommunication to visible wavelengths for G4V compatibility. Optical switches (S) direct photons to designated memory cells. Through controlled photon–spin interactions and subsequent photon detection, the photon qubit is mapped onto a G4V electron spin (dark blue) and then transferred to a nuclear spin (orange). Successful detection heralds photon receipt, which is confirmed via classical communication between nodes.
    \textbf{b)} and \textbf{c)} Optical circuit for the local nucleus-nucleus \textsc{cont} gates in the entanglement distillation and swapping steps, respectively. A single photon source (implemented with a separate G4V~\cite{Knall2022}) generates a photon, which is guided by circulators and optical switches (S). It interacts sequentially with two target G4V electron spins until heralded photon measurements confirm entanglement between them. Nucleus–nucleus \textsc{cnot} gates are then realized using local electron–nuclear gates.
    }
    \label{fig:optical_cir}
\end{figure}

\section{Introduction}

Quantum networks~\cite{Zhang2024, wei_towards_2022, Sukachev2021, ruf_quantum_2021, lodahl_quantum-dot_2017, reiserer_cavity-based_2015} capable of distributing high-fidelity entanglement are essential for applications such as distributed quantum sensing~\cite{kim_distributed_2024, zhang_distributed_2021, guo_distributed_2020, Degen2017}, secure communication~\cite{Ramya2025, Chen2021, gisin_quantum_2007}, and networking of quantum computers~\cite{sunami_scalable_2025, aghaee_rad_scaling_2025}. A major challenge for the construction of extended quantum networks is photon loss in optical fibers: $99$\% of photons are lost over every $100$~km of transmission~\cite{miya_ultimate_1979}. Since quantum states cannot be amplified like classical signals~\cite{wootters_single_1982}, quantum repeaters~\cite{briegel_quantum_1998} are required to extend entanglement over continental distances.  

A variety of hardware platforms have been explored for quantum repeaters, including atomic ensembles~\cite{Sangouard2011}, trapped ions~\cite{Krutyanskiy2023}, solid-state defect centers~\cite{Wang2023}, and semiconductor quantum dots~\cite{Wang2012}. Each offers distinct strengths: quantum dots (QDs) can generate entangled photons at high rates with excellent fidelity~\cite{schimpf_quantum_2021}, while group-IV-vacancy (G4V) centers in diamond provide long-lived, optically addressable spins for efficient and heralded photon storage~\cite{knaut_quantum_2024,harris_coherence_2023}. These complementary capabilities naturally motivate hybrid repeater architectures that combine the best features of different hardware systems.  

Here, we propose and analyze such a hybrid scheme, integrating QDs as entangled photon-pair sources with G4V centers as efficient single-spin memories. A key challenge is the bandwidth mismatch: QDs emit photons with $\sim$10~GHz bandwidths~\cite{Neuwirth2021, zajac_quantum_2025, schimpf_quantum_2021}, while G4Vs have narrow optical transitions of order 100~MHz~\cite{Sukachev2017, bopp_sawfish_2024}. To address this challenge, we use a comprehensive model in \cite{strocka_memory_2025} of the cavity-mediated spin--photon interaction that includes polarization effects, magnetic-field-induced level splitting, and cross-talk between transitions. Having expanded beyond more simplified models~\cite{omlor_entanglement_2024}, our approach enables the identification of practical mitigation strategies and provides a realistic assessment of their effectiveness. In particular, we investigate techniques such as frequency filtering and magnetic-field optimization to determine the conditions under which efficient, high-fidelity photon storage can be achieved. 

Building on this, we evaluate the performance of a full repeater chain with entanglement distillation using an operational protocol based on BB84 quantum key distribution~\cite{shor_simple_2000}. Our results show that a network with thousands of G4V memories across multiple nodes can reach secret-key rates on the order of hundreds to thousands of bits per second over continental distances. This demonstrates the strong potential of this hybrid QD--G4V platform as a path toward scalable long-distance quantum networks.  

The remainder of this paper is organized as follows: Sec.~\ref{sec:overview} introduces the operating principles of the scheme, Sec.~\ref{sec:bwm} presents the photon-cavity-G4V model, Sec.~\ref{sec:chain} evaluates repeater chain performance, and Sec.~\ref{sec:dis} concludes with a discussion of future directions.

\section{Hardware Implementation}\label{sec:overview}

The hybrid repeater chain protocol is illustrated in Fig.~\ref{fig:overview1}. The two end nodes, Alice and Bob, are separated by a large distance. The repeater nodes divide the total distance into multiple segments, each containing a QD emitter in the middle for sending photons to the two neighboring nodes. Each repeater node contains two memory modules, one of each segment. Depending on whether entanglement distillation is applied, the end node contains one or no memory modules. When there is no entanglement distillation, the received photons are directly measured, hence no quantum memories at the end nodes are required. Each memory module contains multiple memory cells, each of which contains a G4V electron spin serving as the communication qubit and a nuclear spin, \textit{e.g.}, $^{13}$C~\cite{Tim2024mapping, beukers_control_2024}, $^{29}$Si~\cite{phone_gate_stas2022robust}, $^{117}$Sn \cite{Parker2023}, or $^{119}$Sn, serving as the storage qubit. Here, the G4V electron spin refers to the collection of electrons in the unsaturated covalent bonds~\cite{gu2025group, Hepp2014SiVthesis}. It is effectively equivalent to a hole with spin-$\tfrac{1}{2}$.

Our two-way repeater chain protocol consists of three key operations: elementary link establishment, entanglement distillation~\cite{rozpedek_optimizing_2018, bratzik_quantum_2013, horodecki_quantum_2009, dur_quantum_1999, bratzik_quantum_2013, distillation_nigmatullin2016minimally}, and entanglement swapping~\cite{yan_survey_2021,Sangouard2011}. Elementary link establishment generates entanglement between neighboring nodes separated by a short distance. Entanglement distillation enhances the fidelity of Bell pairs at the cost of consuming multiple pairs. Two possibilities, entanglement distillation for the elementary links and the end-to-end links, are examined in this work. Entanglement swapping extends entanglement by merging two adjacent links into a longer one. We describe these three operations in detail below.

\subsection{Elementary-Link Establishment}
The operation of elementary link establishment contains four steps: entangled photon pair generation, photon transmission, quantum information transfer, and photon-reception heralding. The concrete protocol is illustrated in Fig.~\ref{fig:optical_cir}a.
 
The quantum dot (QD) generates entangled photon pairs with polarization correlations~\cite{QD_1_reimer2012bright, QD_2_heinze2015quantum, QD_3_muller2014demand, QD_4_strobel2024high, schimpf_quantum_2021, Huber2018}. Photon absorption and emission in the QD correspond to an electron transitioning between the conduction and valence bands. As a result, the four lowest-energy quantum states of the QD are: the ground state $\ket{0}$, the two single-exciton states $\ket{\rm X}$ and $\ket{\rm X'}$ of opposite spin orientations, and the biexciton state $\ket{\rm XX}$, as illustrated in Fig.~\ref{fig:overview2}a. Under suitable laser irradiation, the QD is initialized in the biexciton state $\ket{\rm XX}$. From this state, it can decay via two possible cascaded paths: $\ket{\rm XX} \rightarrow \ket{\rm X} \rightarrow \ket{0}$ and $\ket{\rm XX} \rightarrow \ket{\rm X'} \rightarrow \ket{0}$. The first (second) path produces two vertically (horizontally) polarized photons of different frequencies.

Here, we consider the ideal case where the intermediate states $\ket{\rm X}$ and $\ket{\rm X'}$ are energetically degenerate, making the two decay paths indistinguishable by photon frequencies. This results in the generation of a maximally entangled polarization Bell-state: 
\begin{equation} \ket{\psi_{\rm QD}} = \frac{1}{\sqrt{2}}\left( \ket{\rm HH'} + \ket{\rm VV'} \right), \label{eq:qdstate} 
\end{equation} 
where $\ket{\rm H}$ and $\ket{\rm V}$ denote the horizontally and vertically polarized photon states, respectively, and the presence and absence of the prime symbol denote different frequencies. We assume QD photon wavelengths in the telecom spectral range.

The two emitted photons have different frequencies and can thus be separated using a wavelength-division multiplexer (WDMP) \cite{aoyama_optical_1979} and be directed to the two different nodes of the segment. To mitigate polarization instability during fiber transmission, before transmission, a polarization-to-time-bin converter (PTBC) \cite{kupchak_time-bin--polarization_2017} is used to convert the polarization states $\ket{\rm H}$ and $\ket{\rm V}$ ($\ket{\rm H'}$ and $\ket{\rm V'}$) into early and late time-bin states $\ket{\rm E}$ and $\ket{\rm L}$ ($\ket{\rm E'}$ and $\ket{\rm L'}$), respectively.

After the polarization-to-time-bin conversion, the two photons will be sent to the two nodes adjacent to the QD emitter, respectively. When a photon arrives at a repeater node, a frequency converter~\cite{albrecht_waveguide_2014} shifts its frequency from the telecom band to the frequency of the G4V spin-photon interface in the visible light regime. The photon is then routed to a designated memory cell by an optical switching network. Through a tailored interaction between the photon and the target G4V electron spin~\cite{bhaskar_experimental_2020,reiserer_cavity-based_2015}, the quantum information (qubit) carried by the photon is transferred to the electron spin. This interaction is discussed in detail further down.

Photons experience significant loss during transmission through optical fibers and on-chip circuits. Therefore, heralding their successful arrival is essential for identifying the memory cells that have been successfully entangled and are ready for the subsequent operation of entanglement swapping. At each repeater node, local heralding is achieved by measuring the photon in the $X$-basis, $\ket{\pm} = \frac{1}{\sqrt{2}}(\ket{\rm E} \pm \ket{\rm L})$, after the quantum state transfer operation~\cite{bhaskar_experimental_2020,reiserer_cavity-based_2015}, i.e. the interaction between the photon and the target electron spin. Moreover, a successfully transferred qubit is further transferred to a nuclear spin via an electron-nuclear entanglement gate \cite{beukers_control_2024} for long-time storage. Lastly, to determine whether the other photon of the same pair has also been received at the opposite node, classical communication between the two nodes is required to exchange heralding signals.

\subsection{Entanglement distillation and swapping}
\label{sec:distill_and_swapping}

We examine the possibilities to apply entanglement distillations for the elementary entanglement links and the end-to-end links as shown in Fig.~\ref{fig:overview1}b and e. The distillation protocol we adopted is presented in Ref.~\cite{distillation_nigmatullin2016minimally}. It introduces three levels of purification. By applying single-qubit rotations, different error types in the entangled state are corrected at successive levels. A level-1, -2, and -3 distilled Bell pair requires at least two, four, and six raw Bell pairs, respectively, as shown in Fig.~\ref{fig:overview1}c. The basic operation in this scheme is the fusion of two Bell pairs into one. Firstly, two pairs of remotely entangled qubits are established between two nodes. Then a \textsc{cnot} gate is implemented locally at each node.

A direct nuclear-nuclear \textsc{cnot} gate could be implemented if both spins belong to the same crystal and are coupled to a comon G4V center~\cite{reyes2022completeBell}. However, due to extensive optical multiplexing requirement, we consider the local nuclear-nuclear \textsc{cnot} gates mediated by the two G4V electron spins and additional photons. The reason is as follows. While coupling to tens of $^{13}$C spins has been demonstrated in NV centers~\cite{Tim2024mapping}, high-fidelity operations are typically feasible only with a small subset of strongly coupled spins. Moreover, because G4V centers exhibit long reset times and slower G4V--$^{13}$C gates compared to fast quantum dot emitters, hundreds of G4Vs are required for sufficient multiplexing. Thus, the probability that two targeted spins reside at the same site is low, making inter-G4V operations necessary for scalable architectures. We therefore focus on on a scenario where each nuclear spin is hosted by a separate G4V.

A schematic optical circuit for the fusion operation is shown in Fig.~\ref{fig:optical_cir}b. The procedure consists of two steps. First, entanglement is generated between the two electron spins~\cite{Knaut2024}. This step can be retried until either success or a maximum number of attempts is reached. A cap in the number of attempts ensures synchronization across the repeater chain and limits decoherence in the nuclear memory. In each attempt, a single photon—emitted by a G4V source or another compatible emitter—is prepared as a time-bin qubit and routed through an optical circuit containing two circulators. The photon sequentially interacts with the electron spins of both G4V-memories before being measured in the $X$ basis. Successful detection heralds entanglement between the electron spins. If no heralding event is observed, the G4Vs are re-initialized while the nuclear spin states remain preserved.

Once entanglement between two electron spins is established, the second step performs local Bell-state measurements between each electron and its adjacent nuclear spin. These measurements, enabled by electron--nuclear entangling gates, realize a near-deterministic \textsc{cnot} gate between the two nuclear spins~\cite{beukers_control_2024}.

Before advancing to the second and third levels of distillation, classical communication is used to verify the success of the preceding step. Over long distances, however, the communication delay can cause significant decoherence in the quantum states. For the final distillation between Alice and Bob, we therefore consider a post-selection procedure to minimize the the decoherence errors. In this case, the distillation procedure continues without waiting for confirmation from the other node about the success of the measurement. Since Alice and Bob will ultimately measure their qubits to distill a secret key, they can post-select which of the measurements were successful at the end and thereby minimize the effect of decoherence errors. 

Entanglement swapping is realized by performing a Bell-state measurement (BSM) on two selected nuclear spins within a repeater node, each already entangled with a remote node, as shown in Fig.~\ref{fig:overview1}d. In this setting, the BSM between the two nuclear spins again relies on a \textsc{cnot} gate, implemented by the method described above. The relevant optical circuit is similar and shown in Fig.~\ref{fig:optical_cir}c.

\section{System compatibility}\label{sec:bwm}

\begin{figure}[p]
    \centering
    \includegraphics[width=\columnwidth]{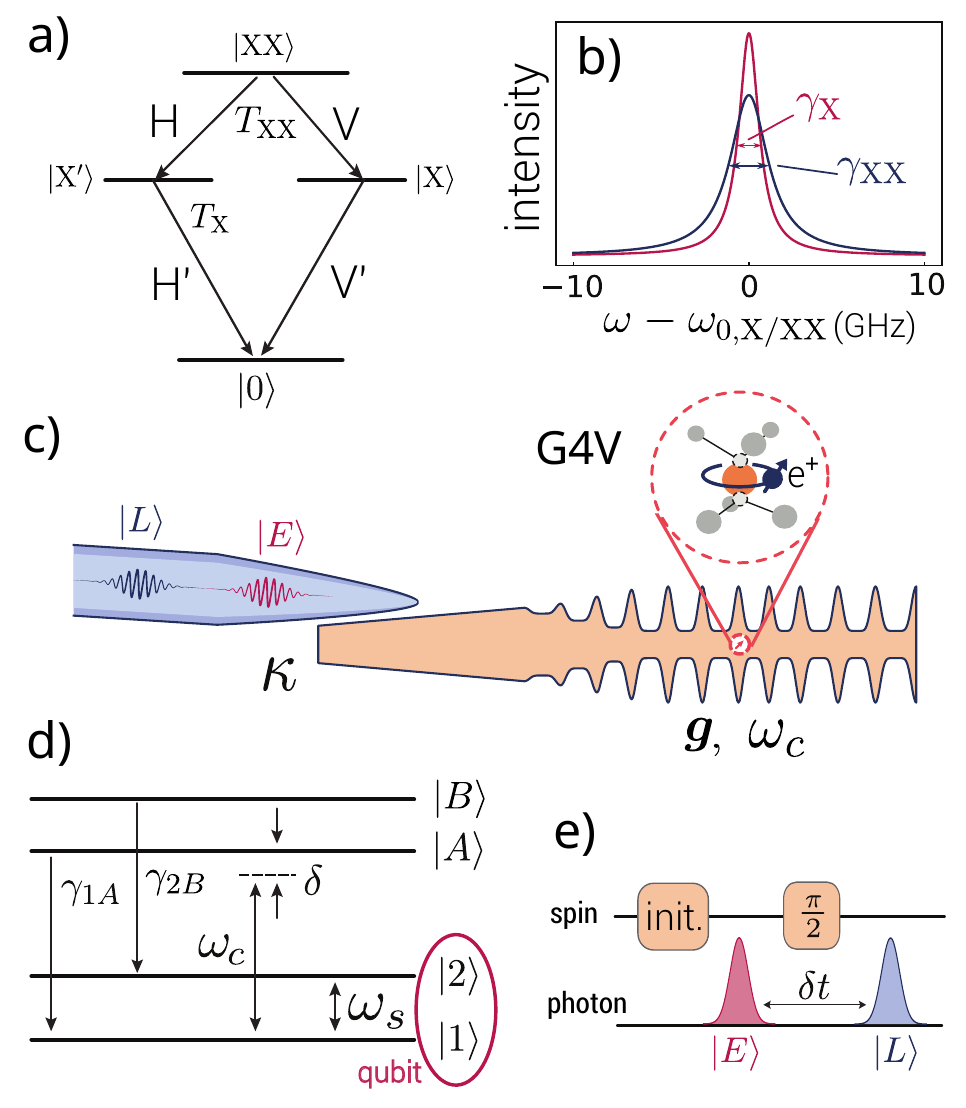}
    \caption{QD and G4V hardware profiles.
    \textbf{a)} Energy levels and decay paths of a quantum dot (QD). The QD includes a ground state $\ket{0}$, two approximately degenerate single-excitation states, $\ket{\rm X}$ and $\ket{\rm X'}$, and a biexcitation state, $\ket{\rm XX}$, which together give rise to two cascade decay paths. The decay times $T_{\rm X}$ and $T_{\rm XX}$ of the single- and biexcitation states correspond to linewidths $\gamma_{\rm X}$ and $\gamma_{\rm XX}$, respectively. Photons are emitted with horizontal (H) or vertical (V) polarization, depending on the decay path; the presence or absence of a prime symbol for the labels H and V denotes different photon frequencies.
    \textbf{b)} Frequency profile of photons emitted by the QD. According to the experiment presented in \cite{schimpf_quantum_2021}, the photons from both $\ket{\rm XX}$, $\ket{\rm X}$ and $\ket{\rm X'}$ exhibit Lorentzian spectral profiles centered at $\omega_{\rm 0,X}$ and $\omega_{\rm 0,XX}$, with bandwidths $\gamma_{\rm XX} = 8.33$~GHz and $\gamma_{\rm X} = 4.34$~GHz, respectively.
    \textbf{c)} Fiber–cavity interface and integration of a group-IV-vacancy color center (G4V) into the sawfish nanophotonic crystal cavity~\cite{bopp_sawfish_2024}. The incoming photon from the fiber is in a superposition of early and late time-bin states, encoding a qubit. The cavity supports a single optical mode with resonance frequency $\omega_c$, coupling strength $\bm{g}=[g_{1A},g_{2B},g_{2A},g_{1B}]$ to the G4V, and a total loss rate $\kappa$.
    \textbf{d)} Energy levels of the G4V center (SiV or SnV), modeled as a four-level system with ground states $\ket{1}$ and $\ket{2}$ and excited states $\ket{A}$ and $\ket{B}$. The optical transitions $\ket{1} \leftrightarrow \ket{A}$ and $\ket{2} \leftrightarrow \ket{B}$ have linewidths $\gamma_{1A}$ and $\gamma_{2B}$, respectively. The cavity resonance $\omega_c$ is detuned by $\delta$ from the $\ket{1} \leftrightarrow \ket{A}$ transition. The ground-states $\ket{1}$ and $\ket{2}$ circled by the orange line make up the qubit of the G4V spin, whose splitting is denoted as $\omega_s$.
    \textbf{e)} Operational sequence for implementing the spin–photon entanglement gate via the reflection-based scheme (see App.~\ref{state_transfer}). The protocol consists of four steps: (1) initialize the G4V spin in state $\ket{1}$; (2) scatter the early time-bin photon; (3) apply a $\pi/2$ rotation around the $y-$axis to the spin; (4) scatter the late time-bin photon. Fig. e) is adapted from \cite{bhaskar_experimental_2020}.}
    \label{fig:overview2}
\end{figure}

\begin{figure}
    \centering
    \includegraphics[width=\linewidth]{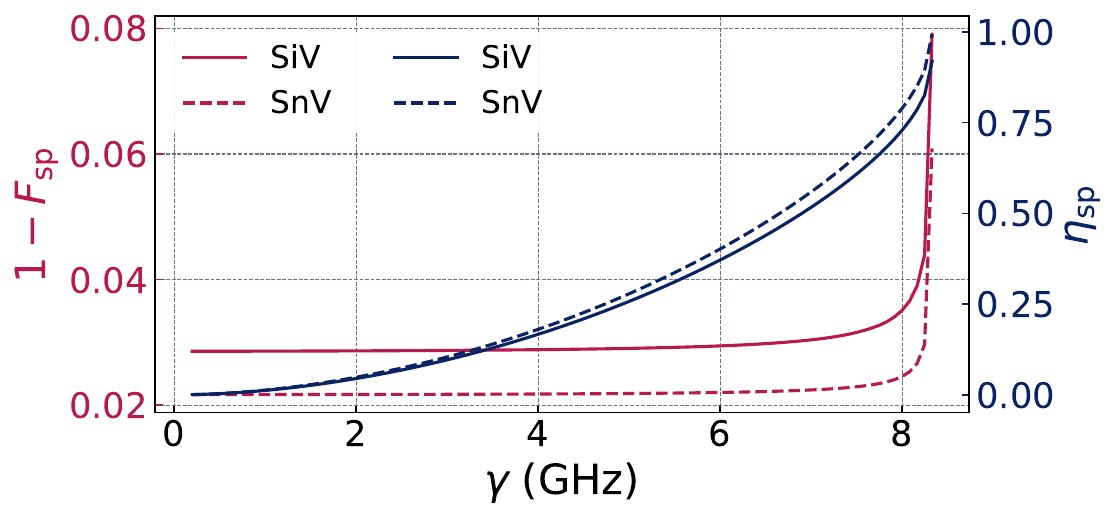}
    \caption{Infidelity $1-F_{\rm sp}$ and efficiency $\eta_{\rm sp}$ of the spin-spin entangled state as a function of the filtered bandwidth $\gamma\coloneqq\tilde{\gamma}_{\rm XX}$ while keeping $\gamma_{\rm X}=4.34$ GHz the same. The relevant parameters for both the SiV and SnV are shown in Tab. \ref{tab:summary} in App. \ref{app:opt}.}
    \label{fig:bw}
\end{figure}

Compatibility between the QD photon emitter and the G4V single-spin quantum memory is a central challenge in our protocol. While the central frequencies of the two systems can be aligned using frequency converters, matching their bandwidths and lineshapes remains a formidable task. The QD emitter has a natural lifetime of $125$–$250$~ps~\cite{schimpf_quantum_2021}, corresponding to a bandwidth on the order of 10~GHz. In contrast, the G4V spin has a natural lifetime of $1.7$–$4.5$~ns~\cite{bopp_sawfish_2024}, implying a bandwidth on the order of 100~MHz—only about 1\% of that of the QD. This mismatch implies that the G4V would fail to interact efficiently with approximately 99\% of the QD-emitted photons. Nevertheless, this issue can be mitigated by coupling the G4V and the broadband photon via an intermediate cavity with an optimized total energy decay rate and central mode frequency relative to the G4V transition \cite{omlor_entanglement_2024}.

Following a decay cascade from the biexcitation state to the ground state, the experimentally investgated QD considered here emits two photons with Lorentzian spectral profiles~\cite{tran_nanodiamonds_2017} and bandwidths $\gamma_{\rm XX} = 8.33$~GHz and $\gamma_{\rm X} = 4.34$~GHz, respectively~\cite{schimpf_quantum_2021}, as shown in Fig.~\ref{fig:overview2}(b). 
Each photon can optionally pass through a Fabry–Perot interferometer for frequency filtering \cite{Ismail2016} (see App. \ref{app:filtering} for details) and is routed to one of the two adjacent nodes containing a G4V spin, thereby distributing entanglement. Since the two photons have different bandwidths, using two distinct cavities optimized for each photon would yield the best performance. However, for simplicity and practical implementation, we adopt a symmetric design for both cavities, balancing system complexity and performance. As a result, only three parameters require optimization: the photon central frequency $\omega_0$ (as targeted by the frequency converters), the cavity resonance frequency $\omega_c$, and the cavity loss rate $\kappa$. 

Because the challenge of low success probability of entanglement transfer, mainly caused by photon loss during fiber transmission and local optical circuits, is addressed through multiplexing, quantum state and gate fidelities become the dominant factor limiting repeater performance. Therefore, our objective is to maximize the fidelity of the resulting G4V spin-spin state relative to the raw G4V electron-electron Bell-state, $\ket{\mathrm{Bell}} = \frac{1}{\sqrt{2}}(\ket{11} + \ket{22})$, where $\ket{ij}$ ($i,j\in \{1, 2\}$) is the quantum state where the left(right) G4V electron in an elementary link is in the $\ket{i(j)}$ state. This leads to the objective function \cite{Liang2019} 
\begin{align}\label{eq:F}
    F_{\rm sp}=\bra{\rm Bell} \rho_{\rm sp} \ket{\rm Bell}
\end{align}
where $\rho_{\rm sp}$ denotes the density matrix of the quantum state of the two G4V spins. 

To examine the quantum state transfer from one of the two photons in the pair to the G4V spin, we consider the setup shown in Fig.~\ref{fig:overview2}(c). Via the polarization to time-bin converter, the photonic qubit is encoded in early and late time bins. The photon is guided through a fiber into a single-sided, overcoupled nanophotonic crystal cavity~\cite{bopp_sawfish_2024}, where it interacts with the G4V spin. After interaction, the photon is reflected back into the fiber and directed to a measurement device. In the simplest case spin-photon entanglement succeeds when the phase contrast fulfills $\Delta\phi(\omega):=\phi_1(\omega)-\phi_2(\omega)=\pi$
on a frequency range around the central frequency of the incoming photon where the expressions $\phi_i$ are the phases of the reflected photon when the spin is initialized in $\ket{i}$ for $i=1,2$. The phase contrast $\Delta\phi$ is determined by the optical splitting $\Delta\omega_s$ of the G4V, which corresponds to the energy difference between the optical transitions connecting the excited states $\ket{A}$ and $\ket{B}$, and the ground states $\ket{1}$ and $\ket{2}$.

Four G4V energy levels are relevant for spin-photon interaction: two spin states in the ground state manifold, $\ket{1}$ and $\ket{2}$, and the other two in the excited state manifold, $\ket{A}$ and $\ket{B}$,  as shown in Fig.~\ref{fig:overview2}(d). Their level splittings depend on the applied magnetic field and local strain. The spin qubit is encoded in $\ket{1}$ and $\ket{2}$, and optical transitions occur between the ground- and excited state manifolds. Of the four allowed transitions, those connecting $\ket{1} \leftrightarrow \ket{A}$ and $\ket{2} \leftrightarrow \ket{B}$ dominate due to favorable dipole strengths and selection rules.

We operate in the weak-driving regime, where the product of the photon’s electric field amplitude and the G4V transition dipole moments is much weaker than the spin’s optical decay rates $\{\gamma_{1A}, \gamma_{2B}\}$. In this regime, the photon induces no population transfer but imparts a conditional phase on the spin qubit, implementing a photon-induced phase gate \cite{strocka_memory_2025}.

Since spectrally degenerate photons are resonant with only one of the optical transitions, simply applying identical phase gates to both time bins would not generate entanglement. However, entanglement can still be created by inserting a spin rotation between the two photon arrivals [see Fig.~\ref{fig:overview2}(e)], provided that the spin rotation does not commute with the phase gates~\cite{borregaard_one-way_2020, bhaskar_experimental_2020}.
Here, we consider a $\pi/2$ rotation around the $y-$axis acting on the G4V spin, which can be implemented via a microwave pulse~\cite{pieplow_efficient_2024} with an error rate below $10^{-4}$~\cite{karapatzakis_microwave_2024}. Following a projective measurement of the photonic qubit, the photonic state is effectively transferred to the spin. Details are provided in App.~\ref{state_transfer}.

When evaluating the objective function in Eq.~\eqref{eq:F}, we account for the broadband nature of the QD photons. A detailed analysis of the associated effects is presented in App.~\ref{app:no_cross}. To enhance the frequency contrast between the $\ket{1} \leftrightarrow \ket{A}$ and $\ket{2} \leftrightarrow \ket{B}$ transitions, we apply an off-axis 1~T magnetic field to both the SiV and SnV centers, which we investigate as light and heavy representatives of G4V, respectively. In addition, a compressive strain of $E_x = 5.38 \cdot 10^{-5}$, perpendicular to the G4V symmetry axis, is applied to the SiV but not the SnV as an outcome of parameter optimization~\cite{pieplow_efficient_2024} (see App.~\ref{app:contrast} for details). As the angle between the magnetic field and the symmetry axis increases, the spectral contrast between the intended transitions improves, leading to better performance. However, this also increases cross-coupling, the unintended coupling between transitions $\ket{1}\leftrightarrow\ket{B},\ket{2}\leftrightarrow\ket{A}$, 
which introduces nonlinearities in the system’s equations of motion. These are included in our modeling and additionally utilized to minimize entanglement infidelities (see App. \ref{app:opt}). Such cross-coupling terms that have not been addressed yet for G4Vs. 

We determine optimal magnetic field orientations for both SiV and SnV subject to the given conditions and utilize crosstalk to minimize the spin-spin entanglement infidelity using state-of-the-art optimization techniques for the rates $\gamma_{\rm X}=4.34$ GHz and $\gamma_{\rm XX}=8.33$ GHz (see App.~\ref{app:opt}). Assuming a perfect photonic Bell state and no filtering, the resulting minimum infidelity and success probability for the SiV are $1-F_{\rm sp, SiV}=7.91\cdot 10^{-2}$ and $\eta_{\rm sp, SiV}=0.9187$, achieved by a SiV-cooperativity $C_{\rm SiV}=\left.g_{1A}^2/(2\gamma_{1A}\, \kappa)\right\vert_{\rm SiV}\approx 11$ (see App.~\ref{app:opt} for the optimized parameters). However, the spin splitting for the considered magnetic field configuration is $\omega_s\approx 26$ GHz, which is challenging to address using microwave spin control. To address realistic parameters, we also consider a magnetic field strength that produces a spin splitting closer to the experimentally achieved value \cite{karapatzakis_microwave_2024}. We choose $B=0.3$ T yielding $\omega_s\approx 8$ GHz. One limitation of the reduced magnetic field strength is a reduction of the maximum feasible bandwidth of the QD photons. 

To investigate the influence of the photon bandwidths, we use the achievable $1 - F_{\rm sp, SiV}=7.91\cdot 10^{-2}$ as reference value and optimize for bandwidth requirements. A quantum dot pair source with photon bandwidths $\gamma<1.56$ GHz is necessary to achieve the infidelity $1 - F_{\rm sp, SiV}<7.91\cdot 10^{-2}$. Instead of reducing bandwidth and repetition rate, we now introduce filtering of the fast yet broadband QD with $\gamma_{\rm X}=4.34$ GHz and $\gamma_{\rm XX}=8.33$ GHz. We find that filtering boosts performance for reduced magnetic fields. For bandwidths $\gamma_X < 4$ GHz and $\gamma_{XX} < 7.3$ GHz the high magnetic field value for the infidelity of $1 - F_{\rm sp, SiV}=7.91\cdot 10^{-2}$ can be achieved. This is, however, achieved on a reduced overall efficiency $\eta_{\rm sp}<0.33$ which is lower than in the unfiltered case. We conclude that broadband QDs including filtering are compatible with SiV cavities and experimental magnetic field constraints, enabling high-fidelity spin $\pi/2$ microwave rotations (see App. \ref{app:filtering} for details). Moreover, as we will show later, keeping a high fidelity on cost of the efficiency is advantageous for the repeater performance.

For the SnV, we obtain $1 - F_{\rm sp, SnV} = 6.09 \cdot 10^{-2}$, $\eta_{\rm sp, SnV} = 0.9935$, and $C_{\rm SnV} = 160$. A higher cooperativity $C_{\rm SnV}$ compared to $C_{\rm SiV}$ is required due to the much smaller natural atomic decay rate $\gamma_{1A,\mathrm{SnV}}$. Since such a high cooperativity for the SnV has not yet been experimentally demonstrated, we evaluate the case where $C_{\rm SnV} = 25$, matching already experimentally demonstrated values \cite{bhaskar_experimental_2020}. For this relatively low cooperativity, a bandwidth as low as 
$\gamma<480$ MHz is required to achieve an infidelity similar to the SiV's spin-spin entangled state infidelity $1-F_{\rm sp}<7.9\cdot 10^{-2}$. 
We conclude that broadband photons, with rates comparable to those of the quantum dot considered here, are compatible with future SnV cavities with cooperativities on the order of $100$. Achieving such high cooperativities may be enabled by advances in high-purity diamond growth \cite{Chakraborty2019}, isotopic purification \cite{Teraji2015}, and optimized surface termination \cite{Hauf2011}. Current state-of-the-art SnV cavities can already interface with broadband QDs through bandwidth compression \cite{Allgaier2017} or spectral filtering. Alternatively, one could employ a QD with intrinsically narrower bandwidth.

Next, we examine the infidelity $1-F_{\rm sp}$ and efficiency $\eta_{\rm sp}$ as a function of $\gamma_{XX}$ fixing $\gamma_{X}=4.34$ GHz for a given magnetic and strain field induced optical splitting using frequency filtering, which we show in Fig. \ref{fig:bw}. The infidelity increases slightly with bandwidth and then rises sharply at larger bandwidths. This behavior reflects the finite contrast of the system; since the cavity and frequency of the incoming photons were previously optimized based on the QD properties, the steep increase occurs near that range. Below we go more into detail of that behavior. As the bandwidth increases, the phase contrast between $\ket{1}\leftrightarrow\ket{A}$ and $\ket{2}\leftrightarrow\ket{B}$ is washed out as was explained in \cite{omlor_entanglement_2024}. The phase contrast can be recovered by increasing the optical splitting $\Delta\omega_s$, which depends on the magnetic field strength, orientation, and strain configuration. Interestingly, the cooperativity cannot be increased indefinitely at a given splitting for increasing the infidelity. This arises because the phase contrast results from a competition between the contrast itself and its first-order derivative with respect to the cooperativity \cite{omlor_entanglement_2024}. As higher cooperativities reduce the magnitude of the first-order derivative, the overall fidelity increases at the price of a reduced phase difference. 

Having discussed the impact of bandwidth and contrast, we next consider the influence of the photon’s temporal profile. Here, the infidelity rise remains small, owing to the spectral shape of the incoming mode. That is of higher order due to spectral filtering. As a consequence, the tails of the spectrum approach zero faster than a Lorentzian. Within the considered bandwidth range the filtered mode remains within the range where $\Delta\phi(\omega)\approx\pi$ and the infidelity has only a slight increase.
However, the more the bandwidth of the filtered mode approaches the QD bandwidth the broader the filter gets. The advantageous effect of the higher order spectral mode gets washed out and the infidelity steeply increases. 
We conclude from our observation that temporal mode shaping without reducing efficiency would be a powerful tool to significantly reduce infidelities.

In contrast to the infidelity, the efficiency rises monotonically as the bandwidth increases. That is due to the increasing total throughput of the filter across all frequencies as the bandwidth increases. It is important to mention that the infidelity for the SnV at $\gamma_X=4.34$ GHz, $\gamma_{XX}=6$ GHz is $1-F_{\rm sp}=0.05$ when no filtering is applied while the infidelity $1-F_{\rm sp}=0.022$ is reached at the mentioned bandwidth when we apply a filter to reduce the bandwidth from $\gamma_{XX}=8.33$ GHz to $\gamma_{XX}=6$ GHz. The vast infidelity difference is due to the incoming intensity spectrum being fourth order instead of second order without a filter. 

Lastly, we analyze the spin-spin entanglement fidelity at the optimized cavity and magnetic field configuration in a local environment of the optimized cavity parameters. For a deviation of maximally $1$ GHz for the cavity loss rate $\kappa$ and cavity mode detuning $\delta:=\omega_{1A}-\omega_c$ the spin-spin entanglement infidelity for the SiV (SnV) stays below $8.11\cdot 10^{-2}$ $(10.0\cdot 10^{-2})$ and efficiency above $0.9175$ $(0.9931)$ (see App. \ref{app:sens} for details). For this particular example, we find that the SiV’s spin–spin entangled state is more robust to fabrication uncertainties than the SnV’s. This does not need to be generally true, but only for this set of local minima of the optimization. We conclude that the spin-spin entanglement fidelity at the optimized cavity and magnetic field configuration is also robust to fabrication uncertainties.

\section{Repeater chain Protocol}\label{sec:chain}

\begin{figure}[t]
\centering
 \includegraphics[width=\linewidth]{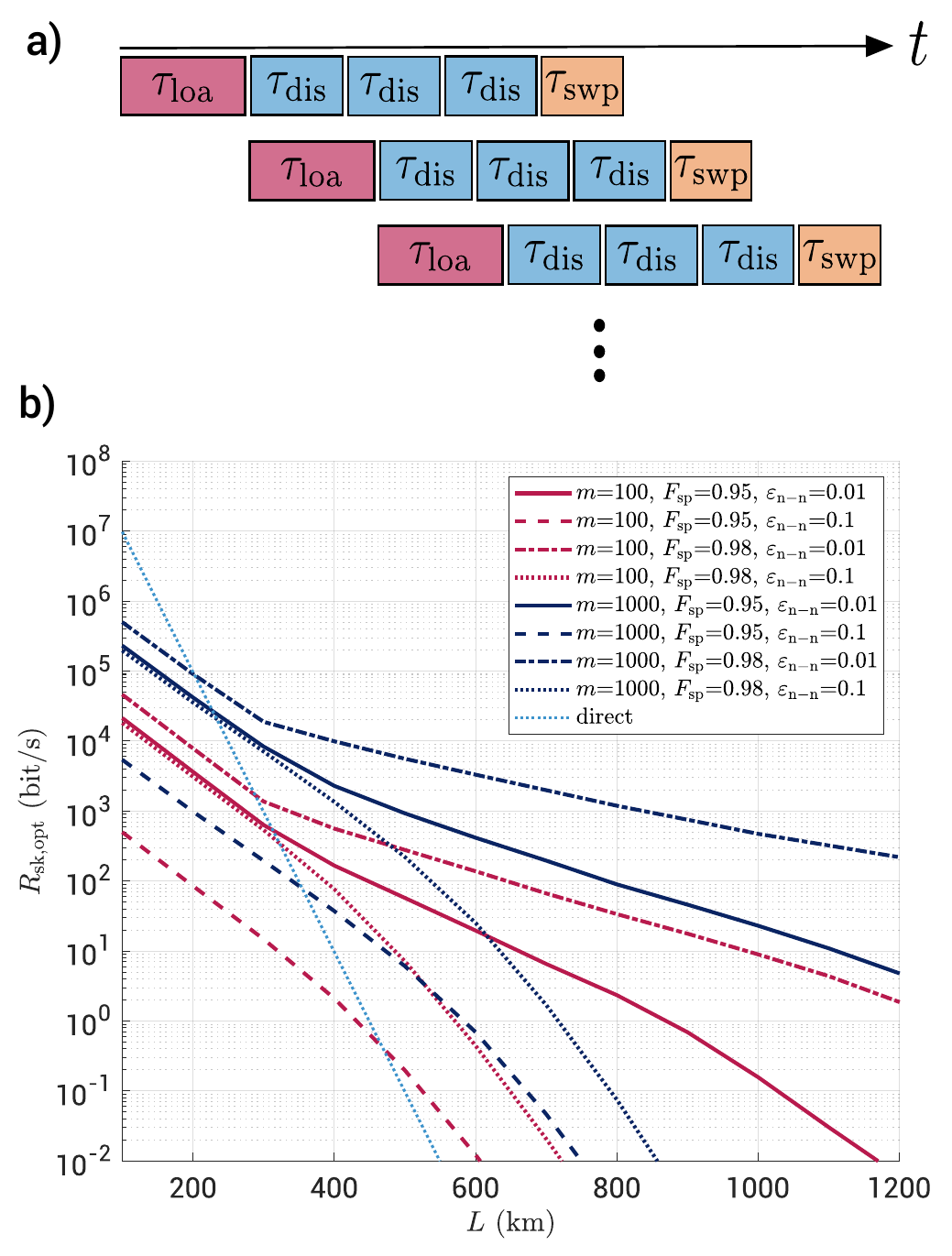}
\caption{Operational protocol and optimal secret-key rate for a QD-SiV quantum repeater chain. 
\textbf{a)} Operation timeline of the repeater chain. The red, blue, and yellow boxes represent the qubit-loading phase, one round of entanglement-distillation operation, and the entanglement swapping phase. Each chain of boxes represents an operational thread of a memory cell.
\textbf{b)} Maximal secret-key rate ($R_{\rm sk, opt}$) as a function of total distance $L$ under certain memory cell number per module ($m$), QD photon bandwidth presented as spin Bell pair fidelity ($F_{\rm sp}$), and the photon-mediated nuclear-nuclear gate error rate $\varepsilon_{\rm n-n}$. The spin Bell pair fidelities $F_{\rm sp}=0.95$ and $0.98$ correspond to the QD photon bandwidth after the frequency filtering of $[\tilde{\gamma}_{\rm X}, \tilde{\gamma}_{\rm XX}]=[4.34, 8.17]$ GHz and $[\tilde{\gamma}_{\rm X}, \tilde{\gamma}_{\rm XX}]=[4.33, 6.50]~\text{GHz}$, respectively. Note that for the $F_{\rm sp}=0.95$ case, only one photon in a Bell pair is filtered ($\tilde{\gamma}_{\rm X}=\gamma_{\rm X}=4.34~\text{GHz}$). The optimal values of  $N$, $n_{\rm loa}$, $n_{\rm dis,n}$, and $n_{\rm dis,e}$, which give the maximal secret-key rate, can be found in App.~\ref{app:opt_m_N}. }
\label{fig:perfo}
\end{figure}

\begin{table}[]
    \centering
    \caption{System parameters assumed in simulations of the QD-SiV hybrid repeater chain.}
    \begin{tabular}{|c|c|}
        \hline
        QD emitter duty cycle ($t_{\rm QD}$) & 1 ns~\cite{QD_4_strobel2024high, schimpf_quantum_2021, Huber2018}  \\
        \hline
        nuclear spin coherence time ($T_{nu}$) & 0.1 s~\cite{SiV_readout1_knaut2024entanglement, SiV_readout2_wei2024universal, SiV_readout3_stas2022robust} \\
        \hline
        fiber attenuation rate ($\gamma_{\rm fib}$) & 0.2 dB/km~\cite{miya_ultimate_1979} \\
        \hline
        classical-signal transmitting speed ($c$) & $2\times 10^5$ km/s \\
        \hline
        \makecell{number of memory cells per \\ memory module ($m$)} & 100, 1000 \\
        \hline
        G4V electron spin reset time ($t_{\rm res}$) & 1 $\mu$s~\cite{bopp_sawfish_2024} \\
        \hline
        electron-nuclear gate time ($t_{\rm nu}$) & 10 $\mu$s~\cite{SiV_readout1_knaut2024entanglement, SiV_readout2_wei2024universal, SiV_readout3_stas2022robust} \\
        \hline
        \makecell{chance of causing decoherence of\\the nuclear spin for each\\electron-electron entanglement\\establishing attempt ($\varepsilon_{\rm nu}$)} & $5\times 10^{-5}$~\cite{SiV_readout1_knaut2024entanglement, SiV_readout2_wei2024universal, SiV_readout3_stas2022robust}\\
        \hline
        \makecell{QD Photon Bell pair fidelity\\ after WDMP and PTBC ($F_{\rm ph}$)} & 0.99~\cite{Huber2018} \\
        \hline
        QD photon bandwidths ($\{\tilde{\gamma}_{\rm X}, \tilde{\gamma}_{\rm XX}\}$) & \makecell{$\{4.34, 8.17\}$ GHz~\cite{schimpf_quantum_2021}\\ $\{4.33,6.50\}$ GHz}\\
        \hline
        \makecell{Spin Bell pair fidelity ($F_{\rm sp}$)} & 0.95, 0.98 \\
        \hline
        \makecell{nucleus-nucleus gate\\error rate ($\varepsilon_{\rm n-n}$)} & 0.01, 0.1 \\
        \hline
        \makecell{maximal number of attempt to\\establish electron-electron\\ entanglement for both entanglement\\ distillation and swapping ($n_{\rm e-e}$)} & 32 \\
        \hline
        cavity-fiber coupling efficiency ($\eta_{\rm c-f}$) & 0.864~\cite{bhaskar_experimental_2020}\\
        \hline
        \makecell{quantum dot photon emitter \\ efficiency ($\eta_{\rm em,qd}$)} & 0.974~\cite{bopp_sawfish_2024}\\
        \hline
        \makecell{group-IV-vacancy color center\\photon emitter efficiency ($\eta_{\rm em,g4v}$)} &0.98~\cite{bopp_sawfish_2024} \\
        \hline
        frequency converter efficiency ($\eta_{\rm fc}$) & 0.73~\cite{Zaske2011}\\
        \hline
        photon detection efficiency ($\eta_{\rm pd}$) & 0.99~\cite{bhaskar_experimental_2020}\\
        \hline
        \makecell{circulator port-1-to-2/2-to-3 \\ efficiency ($\eta_{\rm cir12/cir23}$)} & 0.83~\cite{thorlabs} \\
        \hline
        optical switch efficiency ($\eta_{\rm swi}$)& 0.95~\cite{Zeng2024} \\
        \hline
    \end{tabular}
    \label{tab:simu_para}
\end{table}

To assess the repeater chain performance, we consider a symmetric node configuration and a cyclical operational protocol. This configuration and protocol may not be optimal, but they provide a lower bound on the secret-key rate. Specifically, we consider $N-1$ repeater nodes that evenly divide the total communication distance $L$ into $N$ segments. Each repeater node contains two quantum memory modules, used to establish elementary entanglement links with adjacent segments (see Fig.~\ref{fig:overview1}). The two end nodes have either one or zero memory modules, depending on whether entanglement distillation is applied. When no entanglement distillation is applied, the incoming photons are measured directly, so no quantum memories are required. Each memory module consists of $m$ memory cells, with each cell composed of a G4V electron spin and a nearby nuclear spin. The nuclear spin serves as long-term storage for a single qubit, while the electron spin provides the interface to photonic qubits. Here, we consider $m\in\{100, 1{,}000\}$. 

Based on the working principle introduced in Sec.~\ref{sec:overview}, a memory cell undergoes three operational phases: qubit loading, entanglement distillation, and entanglement swapping. In each memory module, $m_{\rm loa}$ cells are always dedicated to the qubit-loading phase, while the remaining $m_{\rm pro}=m-m_{\rm loa}$ cells are engaged in either entanglement distillation or entanglement swapping.

The qubit-loading phase contains three steps: cell initialization, photon registering, and classical communication. In the initialization step, the G4V electron spin is reset to a specific quantum state, which takes $t_{\rm res}=1\,\mu\text{s}$. In the photon-registering step, qubits carried by photons emitted from the QD source are transferred to electron spins. This step has a fixed period $T_{\rm str}$. During this period, the memory modules continuously receive photons from the QD emitter, and $m_{\rm loa}$ cells in each module attempt to load the qubits. Because of the high emission rate of the QD source, the photon stream is distributed quasi-parallelly across the targeted electron spins. Due to transmission loss in fibers and local optics, many photons are lost before reaching the spins. If an electron spin fails to register a photon, it is reset for the next attempt. This process is repeated $n_{\rm loa}$ times, so that each targeted electron spin has $n_{\rm loa}$ opportunities to register a photon. Consequently, the photon-registering step has a duration
\begin{equation}
    T_{\rm str} = (n_{\rm loa} - 1)\, \max(t_{\rm res}, m_{\rm loa} t_{\rm QD}) + m_{\rm loa} t_{\rm QD},
    \label{eq:t_loa}
\end{equation}
where $t_{\rm QD} = 1 \,\text{ns}$~\cite{QD_4_strobel2024high, schimpf_quantum_2021, Huber2018} is the duty cycle of the QD emitter.

Once a photon is successfully registered, the cell immediately enters the classical-communication step. Two processes occur simultaneously. First, a classical signal is sent to the opposite node of the segment to announce the success, which takes
\begin{equation}
    t_{\rm com} = \frac{L}{cN},
\end{equation}
where $c=2\times10^5$ km/s is the transmission speed of a classical signal. At the same time, the qubit is transferred from the electron spin to the nuclear spin, which has a longer coherence time $T_{\rm nu}=0.1\,\text{s}$~\cite{SiV_readout1_knaut2024entanglement, SiV_readout2_wei2024universal, SiV_readout3_stas2022robust}. The electron-nuclear transfer requires $t_{\rm nu}=10\,\mu\text{s}$~\cite{SiV_readout1_knaut2024entanglement, SiV_readout2_wei2024universal, SiV_readout3_stas2022robust}. Therefore, the duration of the qubit-loading phase is
\begin{equation}
    \tau_{\rm loa} = t_{\rm res}+T_{\rm str}+\max(t_{\rm com}, t_{\rm nu}).
\end{equation}

During this phase, the expected number of cells that successfully register a photon is
\begin{equation}
    m_{\rm reg}=m_{\rm loa}\left[1-(1-p_{\rm arm})^{n_{\rm loa}}\right].
\end{equation}
Out of these, only
\begin{equation}
    m_{\rm p}=\text{round}(m_{\rm reg}p_{\rm arm})
\end{equation}
are expected to be successfully entangled with a partner cell in the opposite node of the segment. Here $\text{round}(\cdot)$ denotes rounding to the nearest integer, and
\begin{equation}
    p_{\rm arm} = 10^{-\frac{\gamma_{\rm fib}}{10}\cdot \frac{L}{2N}} \, p_{\rm trn}
\end{equation}
is the probability of loading one qubit into one arm of a segment, where $\gamma_{\rm fib}=0.2$ dB/km~\cite{miya_ultimate_1979} is the fiber attenuation rate, and
\begin{equation}
    p_{\rm trn} = \sqrt{\eta_{\rm em,qd}} \; \eta_{\rm fc} \; \eta_{\rm cir12} \; \eta_{\rm cir23} \; \eta_{\rm c-f}^{2(4)} \; \sqrt{\eta_{\rm sp}} \; \eta_{\rm swi}^2 \; \eta_{\rm pd}
\end{equation}
accounts for all optical transmission and detection efficiencies (see Tab.~\ref{tab:simu_para}). The exponent of $\eta_{\rm c-f}$ is 2 or 4 depending if the Fabry-P\'{e}rot interferometer for frequency filtering is applied or not.

At the end of the qubit-loading phase, the $m_{\rm p}$ entangled cells hold qubits and proceed to entanglement distillation. Meanwhile, under proper scheduling, at least $m_{\rm p}$ additional cells are released from the $m_{\rm pro}$ pool (previously in distillation or swapping). This ensures that each module can again allocate $m_{\rm loa}$ cells for the next qubit-loading round. Consequently, the cycle period of the entire repeater chain is $\tau_{\rm loa}$. The timeline of operations is shown in Fig.~\ref{fig:perfo}a.

As mentioned above, we adopt the entanglement distillation protocol of Ref.~\cite{distillation_nigmatullin2016minimally}. In this protocol, one can distill raw Bell pairs to the first, second, or third level. Level-1, -2, and -3 distillation consume two, four, and six raw Bell pairs, respectively, and require one, two, and three sequential distillation rounds, as shown in Fig.~\ref{fig:overview1}c. Each round involves similar operations: single-qubit rotations on the nuclear spins, intra-module nucleus–nucleus gates, nuclear-spin readout, and cross-module classical communication. 

Similar to the entanglement-swapping procedure described in Sec.~\ref{sec:distill_and_swapping}, we employ two G4V electron spins as the medium to implement nucleus–nucleus gates. The procedure is as follows: first, establish entanglement between the two electron spins; then perform local electron–nucleus gates. The electron–electron entanglement is photon-mediated and attempted $n_{\rm e-e}$ times. Each attempt consumes time $t_{\rm res}$ and carries a probability $\varepsilon_{\rm nu}=5\times10^{-5}$~\cite{SiV_readout1_knaut2024entanglement, SiV_readout2_wei2024universal, SiV_readout3_stas2022robust} of inducing nuclear-spin decoherence. The success probability per attempt is
\begin{equation}
    p_{\rm e-e}=\eta_{\rm em,g4v}\; \eta_{\rm cir12}^2\;\eta_{\rm cir23}^2\; \eta_{\rm c-f}^4\; \eta_{\rm swi}^4\; \eta_{\rm pd}\;,
\end{equation}
where the parameters are explained and listed in Tab.~\ref{tab:simu_para}. We set $n_{\rm e-e}=32$, chosen to maximize the objective function
\begin{equation}
    f_{\rm e-e}=P_{\rm e-e}^2\, P_{\rm nu}^4\, \mathrm{e}^{-\frac{4t_{\rm res}n_{\rm e-e}}{T_{\rm nu}}}.
\end{equation}
Here,
\begin{equation}
    P_{\rm e-e}=\left[1-(1-p_{\rm e-e})^{n_{\rm e-e}}\right]
\end{equation}
is the overall success probability of establishing electron–electron entanglement, and
\begin{equation}
    P_{\rm nu}=(1-\varepsilon_{\rm nu})^{n_{\rm e-e}}
\end{equation}
is the probability that the nuclear spin remains error-free during the $n_{\rm e-e}$ attempts.

A nuclear single-qubit rotation or nuclear-spin readout each requires time $t_{\rm nu}$. Thus, the duration of one distillation round is
\begin{equation}
    \tau_{\rm dis}=3t_{\rm nu}+t_{\rm res}n_{\rm e-e}+t_{\rm com}.
\end{equation}
Once the memory cells complete distillation, they enter the entanglement-swapping phase, which involves one nucleus–nucleus gate and nuclear-spin readout. Its duration is
\begin{equation}
    \tau_{\rm swp}=2t_{\rm nu}+t_{\rm res}n_{\rm e-e}.
\end{equation}

At the end of each cycle, the number of established end-to-end links is determined by the minimum number of elementary links across all segments. The expected number of end-to-end links is (see App.~\ref{App:expected})
\begin{equation}
    \mathrm{E}[n_{\rm end}] = P_{\rm e-e}^{N-1}
    \sum_{l=1}^{m_{\rm s}}\left(\sum_{k=l}^{m_{\rm s}} f_{\rm b}(k, m_{\rm s}, p_{\rm arm}) \right)^N ,
    \label{eq:expe_n_end}
\end{equation}
where $m_{\rm s}$ is the number of Bell pairs distilled from the $m_{\rm p}$ raw pairs, and
\begin{equation}
    f_{\rm b}(k, m_{\rm s}, p_{\rm arm})=\binom{m_{\rm s}}{k} p_{\rm arm}^k(1-p_{\rm arm})^{m_{\rm s}-k}
\end{equation}
is the binomial distribution.

After entanglement swapping, the end-to-end links undergo another round of entanglement distillation. Unlike segment-level distillation, this round suffers from long classical-communication delays between end nodes, which cause significant nuclear-spin decoherence. Therefore, the second and third distillation levels are executed without the classical communication for identifying which Bell pair is errant. Errant pairs are allowed to propagate to higher levels, but are identified and discarded later by post-selection.

Using superoperators, the average density matrix of the final end-to-end Bell pairs is
\begin{equation}
    \rho_{\rm f}=\mathcal{D}_{\rm e}\biggl(\mathcal{S}_{N-1}\Bigl([\mathcal{D}_{\rm n}(\rho_{\rm sp})]^{\otimes N}\Bigr)\biggr),
\end{equation}
where $\rho_{\rm sp}$ is the density matrix of two entangled SiVs in an elementary link (App.~\ref{app:entangle_state}), $\mathcal{D}_{\rm n}(\cdot)$ and $\mathcal{D}_{\rm e}(\cdot)$ denote distillation for elementary links and end-to-end links, respectively, and $\mathcal{S}_{N-1}(\cdot)$ represents $N-1$ entanglement swappings. Each operation includes both intrinsic decoherence and environmental noise. Specifically, each distillation round requires two nucleus–nucleus gates, while each swapping requires one. We examine the cases when these gate error rates take values $\varepsilon_{\rm n-n}\in\{0.1, 0.01\}$.

From $\rho_{\rm f}$ we compute the average qubit error rates (QBERs) for $Z$- and $X$-basis measurements:
\begin{gather}
    Q_{\rm Z}=1-\bra{11}\rho_{\rm f}\ket{11}-\bra{22}\rho_{\rm f}\ket{22},\\
    Q_{\rm X}=1-\bra{++}\rho_{\rm f}\ket{++}-\bra{--}\rho_{\rm f}\ket{--},
\end{gather}
where $\ket{1/2}$ and $\ket{+/-}$ are the $Z$- and $X$-basis states of the end-node qubit. According to the BB84 protocol~\cite{shor_simple_2000}, the secret-key rate per entangled pair is
\begin{equation}
    r_{\rm sk}(\rho_{\rm f})=\max\left(0,\ 1-H(Q_{\rm X})-H(Q_{\rm Z})\right),
\end{equation}
where $H(\cdot)$ is the binary entropy~\cite{noauthor_david_2016}. The ultimate secret-key rate is therefore
\begin{equation}
    R_{\rm sk}=\frac{ r_{\rm sk}\, \eta_{\rm dis,e}\, \mathrm{E}[n_{\rm end}]}{\tau_{\rm loa}},
\end{equation}
where $\eta_{\rm dis,e}$ is the success probability of end-to-end distillation.

All system parameters are summarized in Tab.~\ref{tab:simu_para}. Based on these, we simulate the performance of the QD–SiV hybrid repeater chain. We consider two cases of QD photon bandwidths: $\{\gamma_{\rm X}=4.34\,\text{GHz}, \gamma_{\rm XX}=8.17\,\text{GHz}\}$ and $\{\gamma_{\rm X}=4.33\,\text{GHz}, \gamma_{\rm XX}=6.50\,\text{GHz}\}$. For both cases, the entangled-photon fidelity is assumed to be $0.99$, corresponding to spin–spin fidelities of $F_{\rm sp}=0.95$ and $0.98$, respectively (see Fig.~\ref{fig:bw}). We optimize the secret-key rate $R_{\rm sk,opt}$ by scanning over the engineering parameters $\{N, n_{\rm loa}, n_{\rm dis,n}, n_{\rm dis,e}\}$, where $n_{\rm dis,n}, n_{\rm dis,e}\in\{0,1,2,3\}$ denote the distillation levels for elementary and end-to-end links (level zero meaning no distillation). The simulated maximal $R_{\rm sk,opt}$ as a function of $L$, under different $\{m, F_{\rm sp}, \varepsilon_{\rm n-n}\}$, is shown in Fig.~\ref{fig:perfo}b, with optimal parameters listed in App.~\ref{app:opt_m_N}. As an example, a secret-key rate of $500$ bit/s over $1{,}000$ km is achievable with $m=1{,}000$, $\{\gamma_{\rm X}=4.34\,\text{GHz}, \gamma_{\rm XX}=8.17\,\text{GHz}\}$, $\varepsilon_{\rm n-n}=0.01$, and a division of the total distance into seven segments.

\section{Discussion and Outlook}\label{sec:dis}
In this work, we have studied the performance of a hybrid quantum repeater enabled by negatively charged G4Vs as quantum memories and quantum dots as photon pair sources. Our model incorporates dominant imperfections associated with the system. Within this framework, we optimize the cavity design that is required to realize a controlled-phase gate between a broadband photon emitted from a QD and a narrowband qubit system provided by either the SiV or SnV. These gates enable the transfer of quantum states from entangled photons to the SiV or SnV spins. 

For the reported photon bandwidths $\gamma_{\rm X}=4.34$~GHz and $\gamma_{\rm XX}=8.33$~GHz~\cite{schimpf_quantum_2021}, the spin-spin entanglement infidelities within one section are $7.91\times 10^{-2}$ and $6.09\times 10^{-2}$ for the SiV and SnV, respectively. Using spectral filtering we can additionally interface the G4V's cavity with photons that are narrower than the reported photon bandwidth. For those we find a higher fidelity than without filtering at the same bandwidth due to the steeper shape of the spectrum after the filter compared to a Lorentzian. This shows that temporal photon shaping is an important tool for improving reflective spin-photon interfaces.

We further analyze a specific repeater chain with a chosen node configuration and operational protocol. Taking into account reported efficiencies of quantum-optical devices, nucleus--nucleus gate errors, spin rotation errors, depolarization errors in photonic states, and bandwidth mismatch between QD photons and G4Vs, we derive the maximal secret key rate as a function of the total distance, under constraints set by photon bandwidth, nucleus--nucleus gate errors, and the number of memory cells. Our estimation suggests the feasibility of building a functional repeater chain covering a distance at the order of 1,000~km based on state-of-the-art technology.

Finally, given the versatility of color centers in diamond, we anticipate that our approach is compatible with next-generation repeaters that employ quantum error correction~\cite{borregaard2020one, bopp_sawfish_2024, Ruf2021}.\\

\section*{Acknowledgements}
The authors would like to thank Mohamed Belhassen for his scientific input. F.G. gratefully acknowledges Jeroen Grimbergen for his support with the mathematical aspects of the repeater chain performance assessment. Y.S., G.P. and T.S. acknowledge funding for this project provided by the German Federal Ministry of Education and Research (BMBF, project QPIS, No. 16KISQ032K; project QPIC-1, No. 13N15858, project QR.N, No. 16KIS2185, project EU ERC StG QUREP, No. 851810). J.B. and F.G. acknowledge funding from the NWO Gravitation Program Quantum Software Consortium (Project QSC No. 024.003.037). J.B. acknowledges support from The AWS Quantum Discovery Fund at the Harvard Quantum Initiative.\\

\section*{Author Contributions}
Y.S. made imperfect bandwidth matching related derivations and simulations. F.G. provided analysis for the performance of the repeater chain. G.P., J.B., and T.S. developed the idea and supervised the project.
All authors contributed to writing the manuscript.
\bibliographystyle{aipnum4-1}
\bibliography{arxiv/control.bib}
\cleardoublepage

\onecolumngrid
\begin{appendix}
\section{Quantum State Transfer}
\label{state_transfer}

To demonstrate the principle of quantum state transfer from the photon to a spin~\cite{borregaard_one-way_2020}, we assume an arbitrary photonic state $\alpha\ket{E}+\beta\ket{L}$ and initialize the spin in the $\ket{{1}}$ state. The procedure includes four steps: early-time-bin photon reflection, first $\frac{\pi}{2}$ rotation, late-time-bin photon reflection, and second $\frac{\pi}{2}$ rotation. As a simplified example, we first look at a single-frequency photon, i.e. we assume $\ket{E}=\ket{L}=e^{{\rm i}\omega_0 t}$ and a half-open cavity without photon loss. The evolution of the combined photon-spin state is as follows.
\begin{align}
    &\ket{\psi_0}=\left(\alpha\ket{E}+\beta\ket{L}\right)\ket{{1}}\\
    &\xrightarrow[]{\text{early reflection}} e^{{\rm i}\phi_{1}(\omega_0)}\alpha\ket{E}\ket{{1}}+\beta\ket{L}\ket{{1}}\\
    &\xrightarrow[]{\pi/2 \text{ rotation}} e^{{\rm i}\phi_{1}(\omega_0)}\frac{\alpha}{\sqrt{2}}\ket{E}\left(\ket{{1}}+\ket{{2}}\right)+\frac{\beta}{\sqrt{2}}\ket{L}\left(\ket{{1}}+\ket{{2}}\right)\\
    &\xrightarrow[]{\text{late reflection}} e^{{\rm i}\phi_{1}(\omega_0)}\frac{\alpha}{\sqrt{2}}\ket{E}\left(\ket{{1}}+\ket{{2}}\right)+\frac{\beta}{\sqrt{2}}\ket{L}\left(e^{{\rm i}\phi_{2}(\omega_0)}\ket{{2}}+e^{{\rm i}\phi_{1}(\omega_0)}\ket{{1}}\right)\\
    &\xrightarrow[]{\pi/2 \text{ rotation}} \ket{\psi_{{\rm id}}}= e^{{\rm i}\phi_{1}(\omega_0)}\left(\alpha\ket{E{2}}+\beta\ket{L{1}}\right)\label{eq:ideal_state}
\end{align}
By far, we have realized a maximally entangled state $\ket{\psi_{{\rm id}}}$ between the photon and the spin. Here, we have assume $\phi_{1}(\omega_0)=\phi_{2}(\omega_0)+\pi$, which is the ideal condition. Note that the second $\pi/2$ rotation is just for converting the quantum state into the well-known Bell state. Before that, the photon and spin are already maximally entangled.

The state transfer is fulfilled if the state $\ket{\psi_{{\rm id}}}$ undergoes a measurement for the photon on the $X$-basis, $\ket{\pm}=\frac{1}{\sqrt{2}}(\ket{E}\pm\ket{L})$, followed by single-qubit phase correction for the spin depending on the measurement outcomes. 
\section{Optimizing Spin–Photon CPHASE Gates}
\label{app:no_cross}

To calculate a guess iterate for the controlled phase gate optimization, we neglect cross-talk. To derive the objective function, we first elaborate on the reflection scheme for broadband incoming photons. We model broadband incoming photons in the time domain by
\begin{align}
    a(t)=\epsilon_0 e^{({\rm i}\omega_0-\gamma/2)t}.
\end{align}
The amplitude $\epsilon_0$ is far smaller than the atomic decay rate so that no driving between the ground and excited state is steered. The spectrum is given by
\begin{align}\label{eq:spec}
    S(\omega-\omega_0)=\frac{\epsilon_0}{{\rm i}(\omega-\omega_0)+\gamma/2}.
\end{align}
The corresponding amplitude spectrum is lorentzian \cite{tran_nanodiamonds_2017}.
We perform the reflection scheme for entangled photons with emission rates $\gamma_{\rm X},\gamma_{\rm XX}$ and spectrum $S_{\rm X},S_{\rm XX}$. It is 
\begin{equation}\label{eq:deriv}
\begin{split}
    \ket{\psi}&=(\alpha\ket{EE'}+\beta\ket{LL'})\ket{11}\\
              &=\iint S_{\rm X}(\omega-\omega_0) S_{\rm XX}(\nu-\omega_0)(\alpha\ket{\omega_E \nu_E}+\beta\ket{\omega_L \nu_L})\ket{11}\,{\rm d}\omega\, {\rm d}\nu\\
              &\xrightarrow[]{\text{early reflection}} \iint S_{\rm X}(\omega-\omega_0) S_{\rm XX}(\nu-\omega_0)(\alpha R_1(\omega)R_1(\nu)\ket{\omega_E \nu_E}+\beta\ket{\omega_L \nu_L})\ket{11}\,{\rm d}\omega\, {\rm d}\nu\\
              &\xrightarrow[]{\pi/2\,\,\text{rotation}} \frac{1}{2}\iint S_{\rm X}(\omega-\omega_0) S_{\rm XX}(\nu-\omega_0)(\alpha R_1(\omega)R_1(\nu)\ket{\omega_E \nu_E}+\beta\ket{\omega_L \nu_L})(\ket{11}+\ket{12}+\ket{21}+\ket{22})\,{\rm d}\omega \,{\rm d}\nu\\
              &\xrightarrow[]{\text{late reflection}} \frac{1}{2}\iint S_{\rm X}(\omega-\omega_0) S_{\rm XX}(\nu-\omega_0)(\alpha R_1(\omega)R_1(\nu)\ket{\omega_E \nu_E}\\
              &+\beta\ket{\omega_L \nu_L})(R_1(\omega)R_1(\nu)\ket{11}+R_1(\omega)R_2(\nu)\ket{12}+R_2(\omega)R_1(\nu)\ket{21}+R_2(\omega)R_2(\nu)\ket{22})\,{\rm d}\omega\, {\rm d}\nu
\end{split}
\end{equation}
with $\alpha,\beta\in\mathbb{C}$ such that $\vert\alpha\vert^2+\vert\beta\vert^2=1$.
Assuming that every photon gets detected with unity probability the $X$-measurement has the form \cite{propp}
\begin{align}
    \rho_{++}=\iint {}_{\omega\nu}\langle ++\vert\psi\rangle\langle\psi\vert ++\rangle_{\omega\nu}\,{\rm d}\omega\,{\rm d}\nu,\\
    \rho_{+-}=\iint {}_{\omega\nu}\langle +-\vert\psi\rangle\langle\psi\vert +-\rangle_{\omega\nu}\,{\rm d}\omega\,{\rm d}\nu
\end{align}
with $\ket{+}_\omega=\frac{\ket{\omega_E}+\ket{\omega_L}}{\sqrt{2}}$. We find
\begin{align}
    \langle ++\vert\psi\rangle=\frac{1}{4} S_{\rm X} (\omega-\omega_0)S_{\rm XX}(\nu-\omega_0)\Bigg(\alpha R_1(\omega)R_1(\nu)(\ket{11}+\ket{12}+\ket{21}+\ket{22})\\
    +\beta (R_1(\omega)R_1(\nu)\ket{11}+R_1(\omega)R_2(\nu)\ket{12}+R_2(\omega)R_2(\nu)\ket{21}+R_2(\omega)R_2(\nu)\ket{22})\Bigg),\\
    \langle +-\vert\psi\rangle=\frac{1}{4} S_{\rm X} (\omega-\omega_0)S_{\rm XX}(\nu-\omega_0)\Bigg(\alpha R_1(\omega)R_1(\nu)(\ket{11}+\ket{12}+\ket{21}+\ket{22})\\
    -\beta (R_1(\omega)R_1(\nu)\ket{11}+R_1(\omega)R_2(\nu)\ket{12}+R_2(\omega)R_2(\nu)\ket{21}+R_2(\omega)R_2(\nu)\ket{22})\Bigg).
\end{align}
The resulting states $\rho_{++}$ and $\rho_{+-}$ depend on the integrals
\begin{align}
    &I_{1X}=\frac{1}{4}\int_{\mathbb{R}} \tilde{S}_{\rm X}(\omega-\omega_0) \vert R_1(\omega)\vert^2\,{\rm d}\omega,&\quad &I_{1XX}=\frac{1}{4}\int_{\mathbb{R}} \tilde{S}_{\rm XX}(\omega-\omega_0) \vert R_1(\omega)\vert^2\,{\rm d}\omega,\label{eq:IX1w}\\
    &I_{2X}=\frac{1}{4}\int_{\mathbb{R}} \tilde{S}_{\rm X}(\omega-\omega_0) R_1(\omega)R_2^*(\omega)\,{\rm d}\omega,&\quad &I_{2XX}=\frac{1}{4}\int_{\mathbb{R}} \tilde{S}_{\rm XX}(\omega-\omega_0) R_1(\omega)R_2^*(\omega)\,{\rm d}\omega,\\
    &I_{3X}=\frac{1}{4}\int_{\mathbb{R}} \tilde{S}_{\rm X}(\omega-\omega_0) \vert R_2(\omega)\vert^2\,{\rm d}\omega,& \quad &I_{1XX}=\frac{1}{4}\int_{\mathbb{R}} \tilde{S}_{\rm XX}(\omega-\omega_0) \vert R_2(\omega)\vert^2\,{\rm d}\omega\label{eq:IX3w}
\end{align}
where $\tilde{S}_{k}(\omega-\omega_0)=\mathcal{N}_k S_k(\omega-\omega_0)S^*_k(\omega-\omega_0)$ with normalization constants $\mathcal{N}_k=\gamma_k/(2\pi e_0^2)$ for $k=X,XX$ after measuring out the photon.
The components of the spin-spin states are 
\begin{align}
    &\langle 0\vert\rho_{+\pm}\vert 0\rangle=\vert \alpha\pm\beta\vert^2 I_{1X}I_{1XX},\\
    &\langle 0\vert\rho_{+\pm}\vert 1\rangle=(\alpha\pm\beta)I_{1X}(\alpha^* I_{1XX}+\beta^* I_{2XX}),\\
    &\langle 0\vert\rho_{+\pm}\vert 2\rangle=(\alpha\pm\beta)I_{1XX}(\alpha^* I_{1X}+\beta^* I_{2X}),\\
    &\langle 0\vert\rho_{+\pm}\vert 3\rangle=(\alpha\pm\beta)\alpha^* I_{1X}I_{1XX}+(\alpha\pm\beta)\beta^* I_{2X}I_{2XX},\\
    &\langle 1\vert\rho_{+\pm}\vert 1\rangle=\vert\alpha\vert^2 I_{1X}I_{1XX}\pm\alpha\beta^* I_{1X}I_{2XX}\pm\alpha^*\beta I_{1X}I_{2XX}^*+\vert\beta\vert^2 I_{3XX}I_{1X},\\
    &\langle 2\vert\rho_{+\pm}\vert 2\rangle=\vert\alpha\vert^2 I_{1X}I_{1XX}\pm\alpha\beta^* I_{1XX}I_{2X}\pm\alpha^*\beta I_{1XX}I_{2X}^*+\vert\beta\vert^2 I_{3X}I_{1XX},\\
    &\langle 1\vert\rho_{+\pm}\vert 3\rangle=\vert\alpha\vert^2 I_{1X}I_{1XX}\pm\alpha\beta^* I_{2XX}I_{2X}\pm\alpha^*\beta I_{1X}I_{2XX}^*+\vert\beta\vert^2 I_{3XX}I_{2X},\\
    &\langle 2\vert\rho_{+\pm}\vert 3\rangle=\vert\alpha\vert^2 I_{1X}I_{1XX}\pm\alpha\beta^* I_{2X}I_{2XX}\pm\alpha^*\beta I_{1XX}I_{2X}^*+\vert\beta\vert^2 I_{2XX}I_{3X},\\
    &\langle 1\vert\rho_{+\pm}\vert 2\rangle=\vert\alpha\vert^2 I_{1X}I_{1XX}\pm\alpha\beta^* I_{2X}I_{1XX}\pm\alpha^*\beta I_{1X}I_{2XX}^*+\vert\beta\vert^2 I_{2X}I_{2XX}^*,\\
    &\langle 3\vert\rho_{+\pm}\vert 3\rangle=\vert\alpha\vert^2 I_{1X}I_{1XX}\pm\alpha\beta^* I_{2X}I_{2XX}\pm\alpha^*\beta I_{2X}^*I_{2XX}^*+\vert\beta\vert^2 I_{3X}I_{3XX}^*.
\end{align}
The remaining entries follow from the hermitian property of density matrices.
On each measurement outcome we apply a $\pi/2$ rotation $U_{\pi/2}=R_y(\pi/2)\otimes R_y(\pi/2)$ to produce a known Bell-state, i.e. $\tilde{\rho}_{++}=U_{\pi/2}\rho_{++}U^\dagger_{\pi/2}$ and $\tilde{\rho}_{+-}=U_{\pi/2}\rho_{+-}U^\dagger_{\pi/2}$. 
The total spin state is 
\begin{align}\label{eq:state}
    \tilde{\rho}_{\rm sp}=2\tilde{\rho}_{++}+2 U_{\rm CP}\tilde{\rho}_{+-}U^\dagger_{\rm CP}
\end{align}
with the CPHASE gate $U_{\rm CP}={\rm diag}(1,1,1,-1)$.

Atomic decay limits the efficiency of spin-spin entanglement, i.e. $\eta_{\rm sp}={\rm tr}(\tilde{\rho}_{\rm sp})<1$. To evaluate the fidelity we consider the normalized state $\rho_{\rm sp}=\tilde{\rho}_{\rm sp}/\eta_{\rm sp}$. The ideal state is the Bell-state $\ket{\rm Bell}=1/\sqrt{2}(\ket{11}+\ket{22})$. The fidelity is
\begin{align}\label{eq:fid}
    F_{\rm sp}=\bra{\rm Bell} \rho_{\rm sp} \ket{\rm Bell}
\end{align}
For the repeater chain, we prefer high fidelity rather than success probability to be the objective function because the problem of low efficiency can be efficiently addressed by multiplexing. However, the entangled-state fidelity suffers from the significant nucleus-nucleus gate error rates in entanglement distillation and swapping. We find an optimal cavity configuration $(\omega_c,\kappa)$ and central frequency of the incoming mode $\omega_0$ by solving the optimization problem 
\begin{align}\label{eq:optprob}
    \min 1-F_{\rm sp}(\omega_0,\omega_c,\kappa)
\end{align}
using simplicial homology global optimization \cite{endres_simplicial_2018}. 
In \cite{strocka_memory_2025} we developed a comprehensive spin-photon interaction model including crosstalk.  
To optimize subject to crosstalk, we have to convert the integrals shown in Eqs. \eqref{eq:IX1w}-\eqref{eq:IX3w} to time domain. For that purpose, we define $\mathcal{D}_k$ for $k=1,2$ as the map which maps an input signal $a_{\rm in}$ to the reflected signal $a_{\rm out}$ when the spin is initialized in $\ket{k}$.
It holds
\begin{align}
    &I_{1X}=\frac{\gamma_{\rm X}}{4e_0^2}\int_0^T \vert \mathcal{D}_1 (a_{\rm in,X}(t))\vert^2 \,{\rm d}t,&\quad &I_{1XX}=\frac{\gamma_{\rm XX}}{4e_0^2}\int_0^T \vert \mathcal{D}_1 (a_{\rm in,XX}(t))\vert^2 \,{\rm d}t\label{eq:IX1_},\\
    &I_{2X}=\frac{\gamma_{\rm X}}{4e_0^2}\int_0^T \mathcal{D}_1 (a_{\rm in,X}(t))\mathcal{D}_2^* (a_{\rm in,X}(t))\,{\rm d}t,&\quad &I_{2XX}=\frac{\gamma_{\rm XX}}{4e_0^2}\int_0^T \mathcal{D}_1 (a_{\rm in,XX}(t))\mathcal{D}_2^* (a_{\rm in,XX}(t))\,{\rm d}t\label{eq:IX2_},\\
    &I_{3X}=\frac{\gamma_{\rm X}}{4e_0^2}\int_0^T \vert \mathcal{D}_2 (a_{\rm in,X}(t))\vert^2 \,{\rm d}t,&\quad &I_{3XX}=\frac{\gamma_{\rm XX}}{4e_0^2}\int_0^T \vert \mathcal{D}_2 (a_{\rm in,XX}(t))\vert^2 \,{\rm d}t\label{eq:IX3_}
\end{align}
with a photon-emitting duration $T$. We choose $T$ such that these integrals coincide with the integrals shown in Eqs. \eqref{eq:IX1_}-\eqref{eq:IX3_} when no crosstalk is assumed. The fidelity encountering crosstalk is modeled when evaluating the fidelity shown in Eq. \eqref{eq:fid} using the state from Eq. \eqref{eq:state} with the integrals in the time domain shown in Eqs. \eqref{eq:IX1_}-\eqref{eq:IX3_}.

To optimize the fidelity with crosstalk, we first solve the optimization problem in Eq. \eqref{eq:optprob} without crosstalk. We initialize an optimization method like Nelder-Mead's algorithm \cite{2020SciPy-NMeth} in the respective minimizer found from Eq. \eqref{eq:optprob} and perform the optimization, encountering the time integrals shown in Eqs. \eqref{eq:IX1_}-\eqref{eq:IX3_}.
\section{Spin-Spin Entangled State}
\label{app:entangle_state}
We want to derive a closed integral expression for the spin-spin entangled state modeling imperfect and broadband photon pair generation, spin rotation, and crosstalk. Starting point is the entangled photonic qubit
\begin{align}
    \ket{\psi_0}=\frac{1}{\sqrt{2}}(\ket{EE'}+\ket{LL'}).
\end{align}
To model imperfect photon pair generation, we apply a depolarizing channel \cite{collins_depolarizing-channel_2015}, i.e.
\begin{align}
    \rho=(1-\epsilon)\rho_0+\epsilon I/4
\end{align}
with $\epsilon=4(1-F_{\rm ph})/3$, the fidelity $F_{\rm ph}$ of the photon pair generated by the quantum dot and the pure state $\rho_0=\ket{\psi_0}\bra{\psi_0}$. For simplicity, we employ the notation
\begin{align}
    \rho=\sum \rho_{IJKM}\ket{IJ}\bra{KM}
\end{align}
with $I,K=E,L$ and $J,M=E',L'$. We assume an imperfect spin $\pi/2$ rotation modeled by a map $\Lambda$, i.e.
\begin{align}
    \mathcal{D}_{\pi/2}(\ket{11}\bra{11})=\sum \Lambda^{(mn,kl)} \ket{mn}\bra{kl}.
\end{align}
The reflection scheme entails the early reflection, a spin $\pi/2$ rotation, and a late reflection, and reads 
\begin{align}
    &\sum \rho_{IJKM}\ket{IJ}\bra{KM}11\rangle\bra{11}=\sum \rho_{IJKM} a^I a^J \bar{a}^K \bar{a}^M \ket{11}\bra{11}\\
    \xrightarrow[]{\text{early reflection}} &\sum \rho_{IJKM} (\delta_{IE}\mathcal{D}_1(a^E)+\delta_{IL}a^L)(\delta_{JE'}\mathcal{D}_1(a^{E'})+\delta_{JL'}a^{L'})\\
    &(\delta_{KE}\bar{\mathcal{D}}_1(a^E)+\delta_{KL}\bar{a}^L)(\delta_{ME'}\bar{\mathcal{D}}_1(a^{E'})+\delta_{ML'}\bar{a}^{L'})\ket{11}\bra{11}\\
    \xrightarrow[]{\pi/2\,\,\text{rotation}} &\sum \rho_{IJKM}\sum \Lambda^{(mn,kl)} (\delta_{IE}\mathcal{D}_1(a^E)+\delta_{IL}a^L)(\delta_{JE'}\mathcal{D}_1(a^{E'})+\delta_{JL'}a^{L'})\\
    &(\delta_{KE}\bar{\mathcal{D}}_1(a^E)+\delta_{KL}\bar{a}^L)(\delta_{ME'}\bar{\mathcal{D}}_1(a^{E'})+\delta_{ML'}\bar{a}^{L'})\ket{mn}\bra{kl}\\
    \xrightarrow[]{\text{late reflection}} &\sum \rho_{IJKM}\sum \Lambda^{(mn,kl)} (\delta_{IE}\mathcal{D}_1(a^E)+\delta_{IL}\mathcal{D}_m(a^L))(\delta_{JE'}\mathcal{D}_1(a^{E'})+\delta_{JL'}\mathcal{D}_n {a^{L'}})\\
    &(\delta_{KE}\bar{\mathcal{D}}_1(a^E)+\delta_{KL}\bar{\mathcal{D}}_k(\bar{a}^L))(\delta_{ME'}\bar{\mathcal{D}}_1(a^{E'})+\delta_{ML'}\bar{\mathcal{D}}_l(\bar{a}^{L'}))\ket{mn}\bra{kl}
\end{align}
where $\mathcal{D}_k(a)$ denotes the output mode $a_{\rm out}$ from the time evolution according to the Heisenberg-Langevin equations shown in \cite{strocka_memory_2025} initializing $\sigma_{kk}(0)=1$ for an input mode $a_{\rm in}$.
After measuring out the photon, the state reads
\begin{align}
    \tilde{\rho}_{\rm sp}=(2\tilde{\rho}_{++}+2U_{\rm CP}\tilde{\rho}_{+-}U_{\rm CP})
\end{align}
with $U_{\rm CP}={\rm diag}(1,1,1,-1)$, $\tilde{\rho}_{+\pm}=U_{\pi/2}\rho_{+\pm}U^\dagger_{\pi/2}$
Note that the additional $\pi/2$ rotation on each qubit is only a technical simplification for producing a known Bell-state, but is not necessary.
and
\begin{align}
    \rho_{+\pm}&=\sum_{\substack{I,K\in\{E,L\}, \\ J,M\in\{E',L'\}}} \rho_{IJKM} \sum_{m,n,k,l=1}^2 \Lambda^{(mn,kl)} \zeta_{IJKM}^{mnkl}\ket{mn}\bra{kl}
\end{align}
with 
\begin{align}
    \zeta_{IJKM}^{mnkl}&=\delta_{IE}\delta_{JE'}\delta_{KE}\delta_{ME'}I_{X,1}I_{XX,1}+\delta_{IE}\delta_{JL'}\delta_{KE}\delta_{ML'}I_{X,1}I_{XX,2(n-1)+l}\\
    &+\delta_{IL}\delta_{JE'}\delta_{KL}\delta_{ME'}I_{X,2(m-1)+k}I_{XX,1}+\delta_{IL}\delta_{JL'}\delta_{KL}\delta_{ML'}I_{X,2(m-1)+k}I_{XX,2(n-1)+l}\\
    &\pm\delta_{IE}\delta_{JE'}\delta_{KL}\delta_{ML'} I_{X,k}I_{XX,l}\pm\delta_{IL}\delta_{JL'}\delta_{KE}\delta_{ME'}I_{X,2m-1}I_{XX,2n-1}
\end{align}
where $I_{X,m}$ and $I_{XX,m}$ for $m=1,2,3,4$ denote the components of the vectors $\bm{I}_{X}=(I_{X1},I_{X2},I_{X2}^*,I_{X3})$ and $\bm{I}_{\rm XX}=(I_{XX1},I_{XX2},I_{XX2}^*,I_{XX3})$ with
\begin{align}
    &I_{X1}=\mathcal{N}_{\rm X}\int_0^T \vert \mathcal{D}_1 (a_{\rm in,X}(t))\vert^2 \,{\rm d}t,&\quad &I_{XX1}=\mathcal{N}_{\rm XX}\int_0^T \vert \mathcal{D}_1 (a_{\rm in,XX}(t))\vert^2 \,{\rm d}t\label{eq:IX1},\\
    &I_{X2}=\mathcal{N}_{\rm X}\int_0^T \mathcal{D}_1 (a_{\rm in,X}(t))\mathcal{D}_2^* (a_{\rm in,X}(t))\,{\rm d}t,&\quad &I_{XX2}=\mathcal{N}_{\rm XX}\int_0^T \mathcal{D}_1 (a_{\rm in,XX}(t))\mathcal{D}_2^* (a_{\rm in,XX}(t))\,{\rm d}t\label{eq:IX2},\\
    &I_{X3}=\mathcal{N}_{\rm X}\int_0^T \vert \mathcal{D}_2 (a_{\rm in,X}(t))\vert^2 \,{\rm d}t,&\quad &I_{XX3}=\mathcal{N}_{\rm XX}\int_0^T \vert \mathcal{D}_2 (a_{\rm in,XX}(t))\vert^2 \,{\rm d}t\label{eq:IX3}
\end{align}
with a sufficiently large photon emitting duration $T$, normalization constants $\mathcal{N}_k=\gamma_k/(2e_0^2)$ and $a_{\rm in,k}(t)=e_0 e^{({\rm i}\omega_0-\gamma_k/2)t}$ with $k=X,XX$. We choose $T$ so large that the integrals shown in Eqs. \eqref{eq:IX1}-\eqref{eq:IX3} assuming no crosstalk coincide with the integrals in the frequency domain shown in Eqs. \eqref{eq:IX1w}-\eqref{eq:IX3w}.

The success probability for spin-spin entanglement is $\eta_{\rm sp}={\rm tr}(\tilde{\rho}_{\rm sp})$.
The ideal state is the Bell-state $\sigma=\ket{\rm Bell}\bra{\rm Bell}$ with $\ket{\rm Bell}=1/\sqrt{2}(\ket{11}+\ket{22})$. The fidelity of the spin-spin entangled state is
\begin{align}
    F_{\rm sp}=\bra{\rm Bell} \rho_{\rm sp} \ket{\rm Bell}
\end{align}
with $\rho_{\rm sp}=\tilde{\rho}_{\rm sp}/\eta_{\rm sp}$.

We evaluate the spin-spin entangled state for the photon pair fidelity $F_{\rm ph}=0.99$ assuming imperfect spin gates generated via microwaves with the gate fidelity of $F_{\rm mw}=0.9999$ and the optimized cavity configuration and central frequency of the incoming photon for the SiV shown in Tab. \ref{tab:summary}. We choose the photon-emitting duration $T=10$ ns. We examine an experimental demonstrated case of $\gamma_{\rm X}=4.34$ GHz, $\gamma_{\rm XX}=8.33$ GHz and apply filtering as described in App. \ref{app:filtering} to produce narrower photons. Our calculation shows that the efficiencies for different cases are 
\begin{gather}\label{eq:eff}
    \eta_{\rm sp}=\begin{cases} 0.9187, & \gamma_{\rm X}=4.34~{\rm GHz},\gamma_{\rm XX}=8.33~\text{GHz}\\
    0.7906, & \gamma_{\rm X}=4.34~{\rm GHz},\tilde{\gamma}_{\rm XX}=8.1653~\text{GHz}\\
    0.4167, & \tilde{\gamma}_{\rm X}=4.3267~{\rm GHz},\tilde{\gamma}_{\rm XX}=6.5~\text{GHz}
    \end{cases}.
\end{gather}
It is important to mention that $\tilde{\gamma}$ means that filtering was applied and the bandwidth of the outcoming mode is written in Eq. \eqref{eq:eff}.
The density matrices for the three cases are
\begin{gather}
\rho_{\rm sp}(\gamma_{\rm X}=4.34~\text{GHz}, \gamma_{\rm XX}=8.33~\text{GHz})=\begin{pmatrix}
0.407 & 0.016{\rm i} & 0.004{\rm i} & 0.405+0.020{\rm i} \\
-0.016{\rm i} & 0.034 & 0 & 0.032-0.016{\rm i} \\
-0.004{\rm i} & 0 & 0.019 & 0.016-0.003{\rm i} \\
0.405-0.020{\rm i} & 0.032+0.016{\rm i} & 0.016+0.003{\rm i} & 0.459 \\
\end{pmatrix},\\
\rho_{\rm sp}(\gamma_{\rm X}=4.34~{\rm GHz},\tilde{\gamma}_{\rm XX}=8.1653~\text{GHz})=\begin{pmatrix}
0.466 & 0.002 + 0.012{\rm i} & 0.002 + 0.014{\rm i} & 0.466 + 0.025{\rm i} \\
0.002 - 0.012{\rm i} & 0.022 & 0 & 0.021 - 0.011{\rm i} \\
0.002 - 0.014{\rm i} & 0 & 0.011 & 0.010 - 0.013{\rm i} \\
0.466 - 0.025{\rm i} & 0.021 + 0.011{\rm i} & 0.010 + 0.013{\rm i} & 0.501\\
\end{pmatrix},\\
\rho_{\rm sp}(\tilde{\gamma}_{\rm X}=4.3267~{\rm GHz},\tilde{\gamma}_{\rm XX}=6.5~\text{GHz})=\begin{pmatrix}
0.486 & 0.002 + 0.007{\rm i} & 0.002 + 0.001{\rm i} & 0.487 + 0.008{\rm i} \\
0.002 - 0.007{\rm i} & 0.009 & 0 & 0.007 - 0.007{\rm i} \\
0.002 - 0.001{\rm i} & 0 & 0.004 & 0.003 - 0.001{\rm i} \\
0.487 - 0.008{\rm i} & 0.007 + 0.007{\rm i} & 0.003 + 0.001{\rm i} & 0.500\\
\end{pmatrix}.
\end{gather}

\section{Supporting Data}
\subsection{Contrast and Cross-Talk}\label{app:contrast}
\begin{figure}[tb]
    \centering
    \includegraphics[width=\linewidth]{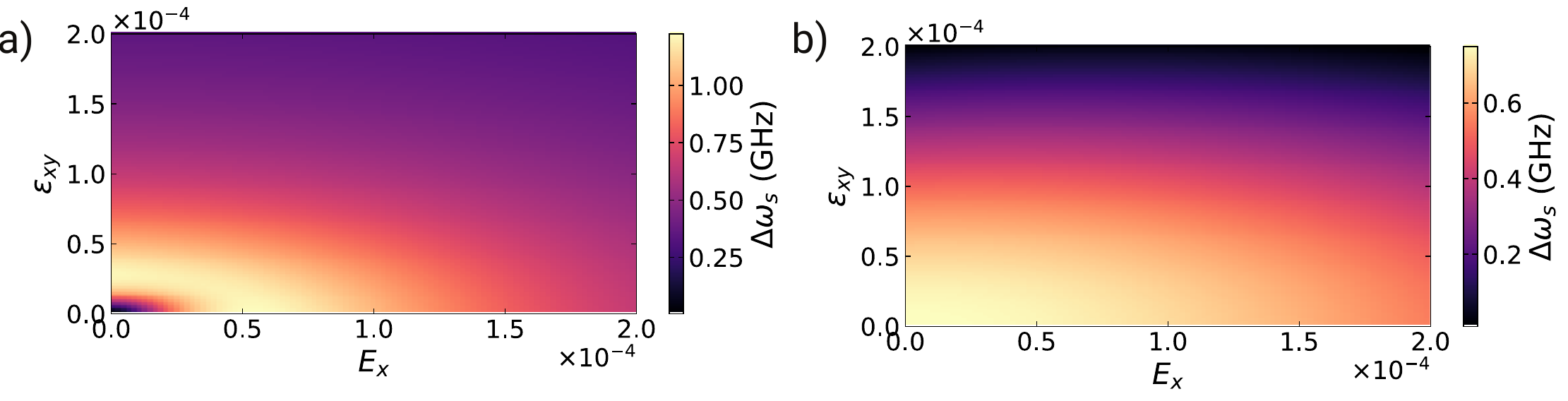}
    \caption{Contrast $\Delta\omega_s:=\omega_{2B}-\omega_{1A}$ as a function of the axial strain $E_x$ and shear strain $\epsilon_{xy}$ for the magnetic field strength $B=1$ T aligned with the symmetry axis. a) Visualization for the SiV and b) SnV. The optimal strain configuration for achieving maximal contrast is $E_x=5.38\cdot 10^{-4}$, $\epsilon_{xy}=0$ for the SiV and $E_x=\epsilon_{xy}=0$ for the SnV.}
    \label{fig:strain}
\end{figure}

The controlled phase gate between the incident photon and the G4V's spin is limited by the contrast $\Delta\omega_s=\omega_{2B}-\omega_{1A}$. The order of magnitude of the contrast depends on strain and the magnetic field. We choose the magnetic field strengths $B=1$ T and $\theta_{\rm dc}=0$ to analyze the contrast's dependence on the extrinsic strain of the SiV and SnV. Fig. \ref{fig:strain} illustrates the contrast $\Delta\omega_s$ as a function of the axial strain $E_x$ and shear strain $\epsilon_{xy}$. The dependence of the G4V's Hamiltonian on strain is shown in \cite{pieplow_efficient_2024}. The maximal contrast for the SiV is roughly $\Delta\omega_s\approx 1$ GHz at $E_x=5.38\cdot 10^{-4}$ and $\epsilon_{xy}=0$. For the SnV, it is $\Delta\omega_s\approx 0.8$ GHz for negligible extrinsic strain. From an experimental perspective, it is not possible to control the extrinsic strain in such a precise manner. We observe a small change in the contrast in the local environment of the optimal strain configuration, making the choice of the optimal strain parameters reasonable for the following simulations.

\begin{figure}[tb]
    \centering
    \includegraphics[width=\linewidth]{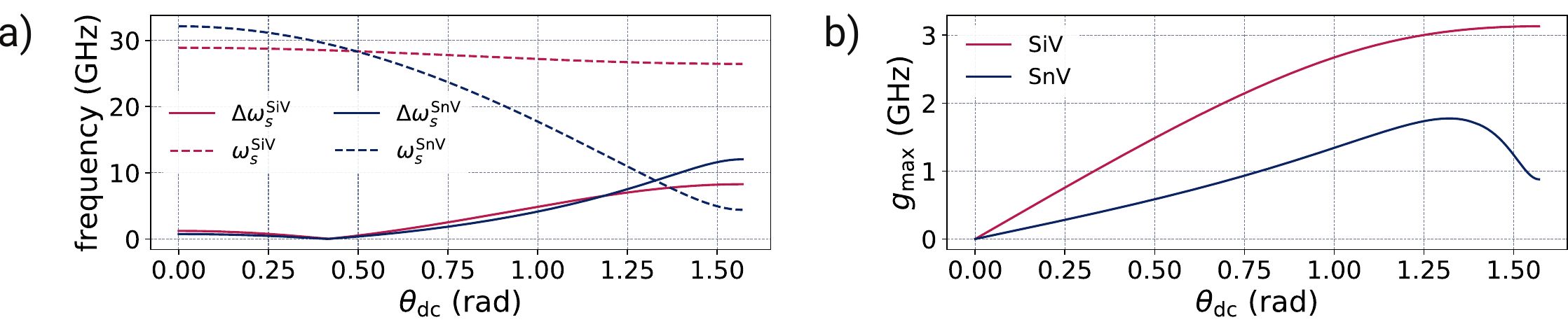}
    \caption{a) Contrast $\Delta\omega_s$ and spin splitting $\omega_s$ as a function of the magnetic field orientation $\theta_{\rm dc}$ for the SiV and SnV. b) Maximal cross coupling strength $g_{\rm max}:=\max\{\vert g_{2A}\vert,\vert g_{1B}\vert\}$ as a function of $\theta_{\rm dc}$.}
    \label{fig:orientation}
\end{figure}

In Fig. \ref{fig:orientation} we visualize the contrast $\Delta\omega_s$, spin splitting $\omega_s$, and the cross couplings $g_{1B},g_{2A}$ as a function of the magnetic field orientation $\theta_{dc}$. On the left-hand side, we observe that the contrast decreases for a value of $\theta_{\rm dc}$ until $\theta_{\rm dc}\approx 0.4$. For $\theta_{\rm dc}>0.4$ the contrast increases monotonically until it reaches the value $\Delta\omega_s\approx 8(12)$ GHz for the SiV (SnV). The spin splitting decreases monotonically for increasing $\theta_{\rm dc}$ for the SiV and SnV. On the right-hand side, it becomes apparent that the cross-talk reaches its maximum at $\theta_{\rm dc}=\pi/2$ $(1.4)$ for the SiV (SnV). From \cite{strocka_memory_2025}, it becomes apparent that the cross-talk spin splitting ratio is a relevant measure to assess the impact of the cross-talk on the spin-photon interaction. That ratio is 
\begin{align}
    r(\theta_{\rm dc})=\frac{\omega_s(\theta_{\rm dc})}{g_{\rm max}(\theta_{\rm dc})},\quad g_{\rm max}:=\max\{\vert g_{2A}\vert,\vert g_{1B}\vert\}.\label{eq:ratio}
\end{align}
When the ratio is small enough, cross-talk influences spin-photon interaction.

\subsection{Optimization}
\label{app:opt}
We assume the QD emission rates $\gamma_{X}=4.34$ GHz, $\gamma_{\rm XX}=8.33$ GHz \cite{schimpf_quantum_2021} at the optimal strain configuration for the SiV and SnV and the magnetic field strength $B=1$ T. Due to the contrast dependence on the magnetic field orientation, we optimize for spin-photon entanglement for $0\le \theta_{\rm dc}\le\pi/2$ for the SiV and SnV to extract an optimal $\theta_{\rm dc}$ for the present configuration.

\begin{figure}[tb]
    \centering
    \includegraphics[width=\linewidth]{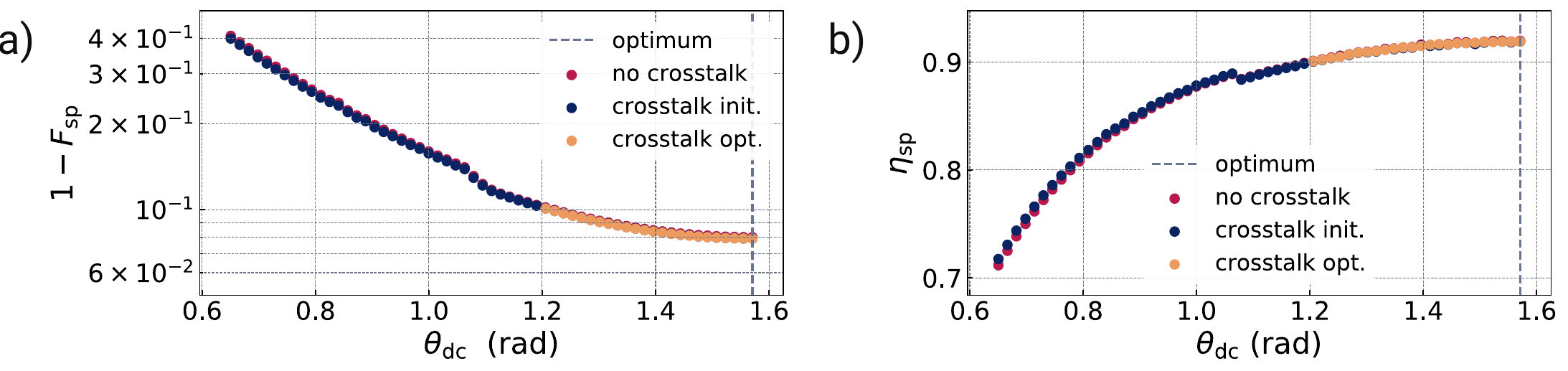}
    \caption{SiV's performance as a function of the magnetic field orientation with and without cross-talk. a) Infidelity $1-F_{\rm sp}$ of the spin-spin entangled state as a function of the magnetic field orientation $\theta_{\rm dc}$ with and without cross-talk. b) Spin-spin entanglement success probability $\eta_{\rm sp}$ as a function of $\theta_{\rm dc}$ with and without cross-talk. The dashed gray line denotes the magnetic field orientation $\theta_{\rm dc}$ where the infidelity is minimal.}
    \label{fig:SiV_theta}
\end{figure}

In Fig. \ref{fig:SiV_theta} we illustrate the performance of spin-photon entanglement as a function of the magnetic field orientation for the SiV. In Fig. \ref{fig:SiV_theta}a) we illustrate the infidelity curve for $\theta_{\rm dc}\ge 0.7$ because the optical splitting $\Delta\omega_s$ is too small to benefit from spin-photon entanglement for $\theta_{\rm dc}<0.7$.
For $\theta_{\rm dc}> 0.7$, the infidelity decreases, where the infidelity with and without cross-talk are so close that no difference becomes apparent. The decreasing infidelity is due to the rising contrast $\Delta\omega_s$. A magnetic field orientation $\theta_{\rm dc}=\pi/2$ produces an optimized infidelity with the value $1-F_{\rm sp}=7.92\cdot 10^{-2}$ including crosstalk. Without crosstalk, the infidelity $1-F_{\rm sp}=8.00\cdot 10^{-2}$ is achieved at the considered magnetic field orientation. That is counterintuitive because crosstalk refers to unwanted couplings, which produce a nonlinear disturbance in the Heisenberg-Langevin equations shown in \cite{strocka_memory_2025}. However, the optical splitting between the cross-transitions $\omega_{1B}-\omega_{2A}$ is larger than the optical splitting between the resonant transitions $\omega_{1A}-\omega_{2B}$. That means that crosstalk produces a gain of optical splitting in a nonlinear manner, resulting in a potentially lower infidelity. For the present simulations, the crosstalk is small enough that its nonlinear behavior, like frequency mixing, a property of nonlinear oscillations \cite{Vanhille2014}, does not deteriorate the infidelity. Local optimization improves the infidelity for the present parameters up to $1-F_{\rm sp}=7.91\cdot 10^{-2}$.
In Fig. \ref{fig:SiV_theta}b), the spin-spin entanglement success probability is illustrated as a function of the magnetic field orientation. The efficiencies with and without cross-talk differ so little that they are virtually indistinguishable. The efficiency rises monotonically for $\theta_{\rm dc}>0.7$ since the contrast increases in that range. At the optimal magnetic field orientation, the efficiency is $\eta_{\rm sp}\approx 0.9187$. 

\begin{figure}[tb]
    \centering
    \includegraphics[width=\linewidth]{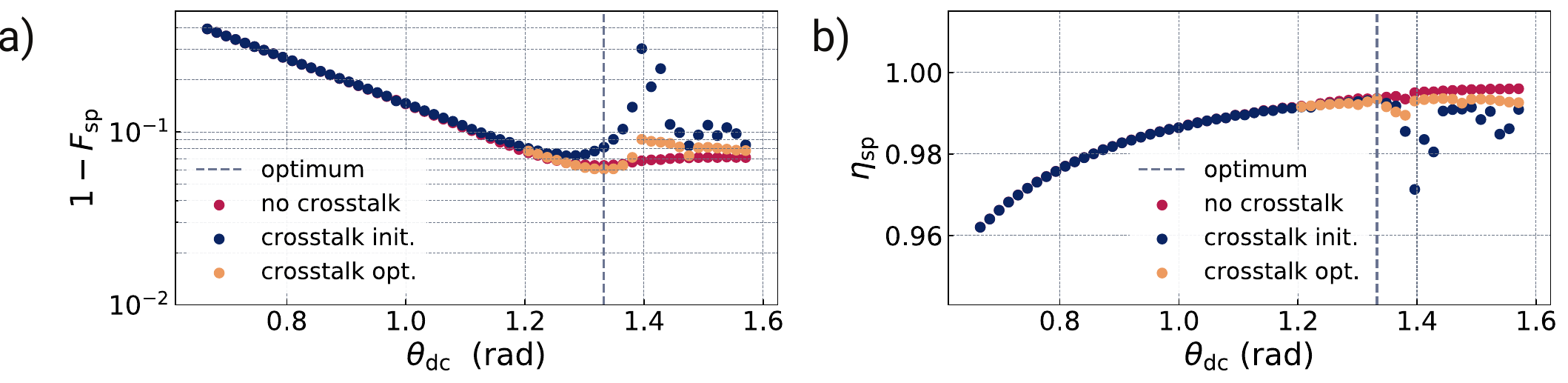}
    \caption{SnV's performance as a function of the magnetic field orientation with and without cross-talk. a) Infidelity $1-F_{\rm sp}$ of the spin-spin entangled state as a function of the magnetic field orientation $\theta_{\rm dc}$ with and without cross-talk. b) Spin-spin entanglement success probability $\eta_{\rm sp}$ as a function of $\theta_{\rm dc}$ with and without cross-talk. The dashed gray line denotes the magnetic field orientation $\theta_{\rm dc}$ where the infidelity is minimal.}
    \label{fig:SnV_theta}
\end{figure}

In Fig. \ref{fig:SnV_theta}, we illustrate the performance of spin-photon entanglement as a function of the magnetic field orientation for the SnV. The behavior of the infidelity on the left-hand side and efficiency on the right-hand side behaves equally compared to the SiV until the optimal magnetic field orientation $\theta_{\rm dc}=1.285$. For $\theta_{\rm dc}>1.285$, the infidelity with cross-talk rises until $\theta_{\rm dc}\approx 1.4$, where the cross couplings also reach their maximum. The infidelity decreases afterwards because the cross couplings decrease for $\theta_{\rm dc}>1.4$. The efficiency in Fig.~\ref{fig:SnV_theta}b) rises monotonically, except near $\theta_{\rm dc}=1.4$. At the orientation $\theta_{\rm dc}=1.33$ we find the infidelity $8.16\cdot 10^{-2}$ with crosstalk and $6.40\cdot 10^{-2}$ without crosstalk. Compared to the SiV crosstalk leads to an increase in the infidelity. However, local optimization using Nelder-Mead's algorithm \cite{2020SciPy-NMeth} is able to improve the infidelity with crosstalk up to $1-F_{\rm sp}=6.09\cdot 10^{-2}$, which is better than the infidelity without crosstalk. That is due to the increase in optical splitting generated by crosstalk. It is not possible to define an optical splitting here because of the nonlinear nature of the Heisenberg-Langevin equations shown in \cite{strocka_memory_2025}. However, it provides an intuitive understanding of the observed phenomenon. We find the optimal infidelity $1-F_{\rm sp}=6.09\cdot 10^{-2}$ and efficiency $\eta_{\rm sp}=0.9935$.

We summarize the optimal $\theta_{\rm dc},\kappa,\delta_c,\delta_0$ with $\delta_0=\omega_{1A}-\omega_0$, $\delta_c=\omega_{1A}-\omega_c$, the respective cooperativities which lead to a specific success rate $\eta_{\rm sp}$ and infidelity $1-F_{\rm sp}$ in Tab. \ref{tab:summary}. We find $1-F_{\rm sp}^{\rm SiV}>1-F_{\rm sp}^{\rm SnV}$ because $\Delta\omega_s^{\rm SnV}>\Delta\omega_s^{\rm SiV}$. Furthermore, we observe $\eta_{\rm sp}^{\rm SnV}>\eta_{\rm sp}^{\rm SiV}$. That is because $\gamma_{1A}^{\rm SnV},\gamma_{2B}^{\rm SnV}<\gamma_{1A}^{\rm SiV},\gamma_{2B}^{\rm SiV}$. Furthermore, it becomes apparent that $C^{\rm SiV}<C^{\rm SnV}$ which is because $\gamma^{\rm SiV}\approx 10\gamma^{\rm SnV}$ for both transitions $1A$ and $2B$. 

Both the optimized configurations for the SiV and SnV have a disadvantage due to experimental restrictions.
The considered magnetic field configuration for the SiV requires microwave spin control to achieve a spin $\pi/2$ rotation for a qubit with the splitting $\omega_s\approx 26$ GHz. That is an experimental challenge. The optimized cooperativity $C_{1A}, C_{2B}\approx 300$ for the SnV is so large that it has not yet been experimentally demonstrated. That represents a disadvantage of the configuration.

To provide data which are closer to state-of-the-art cavities and microwave spin control restrictions we assume the magnetic field $B=0.3$ T and $\theta_{\rm dc}=\pi/2$ for the SiV and find $1-F_{\rm sp}<7.37\cdot 10^{-2}$ for $\gamma<1.6$ GHz at $C_{1A}=25$. For that, we assumed crosstalk, which could improve the infidelity from $1-F_{\rm sp}=8.09\cdot 10^{-2}$ to $1-F_{\rm sp}=7.37\cdot 10^{-2}$ using Nelder-Mead's optimization algorithm \cite{2020SciPy-NMeth}. We conclude that broadband QDs are compatible with SiV cavities and magnetic field constraints, enabling high-fidelity spin $\pi/2$ microwave rotations. 

\begin{figure}[tb]
    \centering
    \includegraphics[width=\linewidth]{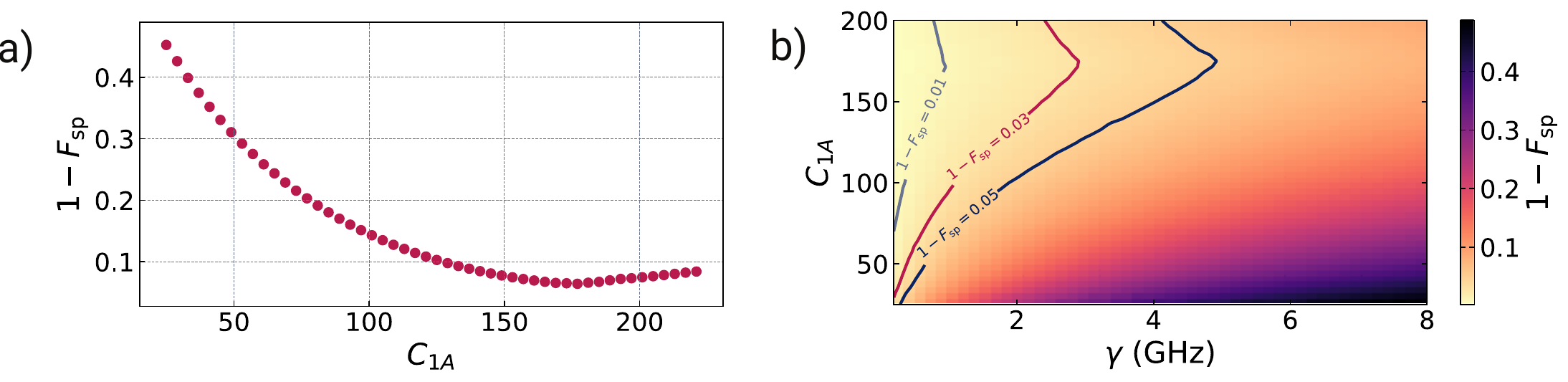}
    \caption{Spin-spin entanglement infidelity depending on the cooperativity. a) Spin-spin entanglement infidelity $1-F_{\rm sp}$ of the SnV as a function of the cooperativity $C_{1A}$. At $C_{1A}, C_{2B}\approx 25$ it is $1-F_{\rm sp}\approx 0.45$ at $\gamma_{\rm X}=4.34$ GHz, $\gamma_{\rm XX}=8.33$ GHz. b) Spin-spin entanglement infidelity $1-F_{\rm sp}$ of the SnV as a function of the cooperativity $C_{1A}$ and bandwidth $\gamma=\gamma_{X,XX}$.}
    \label{fig:performance_C}
\end{figure}

For the SnV, we illustrate the spin-spin entanglement infidelity as a function of the cooperativity at $\theta_{\rm dc}=1.33$. In Fig. \ref{fig:performance_C}a), we visualize that. We observe that the infidelity decreases for an increasing cooperativity. Due to experimental restrictions \cite{bhaskar_experimental_2020} we extract the infidelity at $C_{1A},C_{2B}\approx 25$ and find $1-F_{\rm sp}\approx 0.45$. In Fig. \ref{fig:performance_C}b), we visualize the infidelity as a function of the cooperativity and bandwidth. We observe that for the fidelity $F_{\rm sp}>0.95$ at $C_{1A}=25$, a bandwidth of $\gamma<0.1$ GHz is required. We conclude that higher cooperativities are of great relevance for the SnV. 

\begin{table}[tb]
    \centering
    \caption{Optimized parameters with the respective values for the efficiency $\eta_{\rm sp}$ and infidelity $1-F_{\rm sp}$ for the SiV and SnV with $\delta_c=\omega_{1A}-\omega_c$ and $\delta_0=\omega_{1A}-\omega_0$ for the QD emission rates $\gamma_{\rm X}=4.34$ GHz and $\gamma_{\rm XX}=8.33$ GHz.}
    \begin{tabular}{ccccccccccccc}
       Center  & $\theta_{\rm dc}$ [rad] & $\kappa$ [GHz] & $\delta_c$ [GHz] & $\delta_0$ [GHz] & $\vert g_{1A}\vert$ [GHz] & $\vert g_{2B}\vert$ [GHz] & $\vert g_{2A}\vert$ [GHz] & $\vert g_{1B}\vert$ [GHz] & $C_{1A}$ & $C_{2B}$  & $\eta_{\rm sp}$ & $1-F_{\rm sp}$\\
       \hline
        SiV & $\pi/2$ & $37.70$ & $-4.07$ & $5.44$ & $12.5$ & $13.17$ & $1.72$ & $3.13$ & $11.39$ & $13.11$ & $0.9187$ & $7.91\cdot 10^{-2}$\\
        SnV & $1.33$ & $6.69$ & $-4.50$ & $-4.51$ & $5.17$ & $5.22$ & $1.76$ & $1.77$ & $155.83$ & $161.86$ & $0.9935$ & $6.09\cdot 10^{-2}$
    \end{tabular}
    \label{tab:summary}
\end{table}
\subsection{Sensitivity}\label{app:sens}
\begin{figure}[tb]
    \centering
    \includegraphics[width=\linewidth]{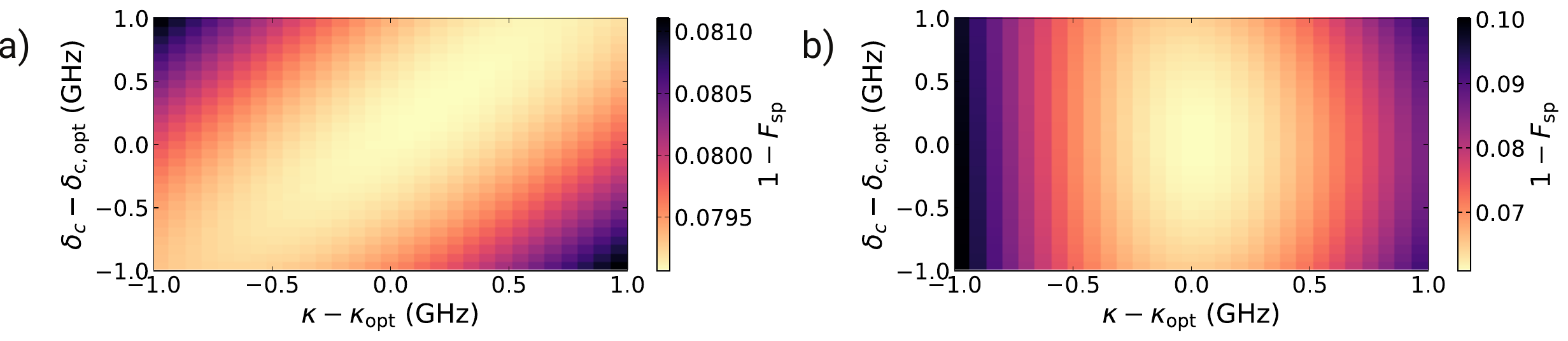}
    \caption{Sensitivity of the spin-spin entanglement infidelity $1-F_{\rm sp}$. a) The infidelity is visualized as a function of the detuning deviation of the cavity mode $\delta_c-\delta_{\rm c,opt}$ and cavity loss deviation $\kappa-\kappa_{\rm opt}$ for the SiV. b) The same visualization is produced for the SnV. The optimal cavity parameters used for the present graph are listed in Tab. \ref{tab:summary}.}
    \label{fig:sensitivity}
\end{figure}

The optimized parameters shown in Tab. \ref{tab:summary} cannot be hit exactly, for example, due to fabrication uncertainties. Therefore, we analyze the dependence of the system's performance on the cavity parameters $\kappa$ and $\delta_c$. We only show the infidelity here since we optimized only for high fidelity spin-spin entanglement. In Fig. \ref{fig:sensitivity} we visualize the infidelity as a function of the cavity parameters for the SiV and SnV. We find that the infidelity changes less in the considered region for the SiV than SnV. That is due to the local nature of the optimum and does not generally have to be true. Based on the illustrations, we conclude that the SiV and SnV's spin-spin entanglement infidelity is robust to fabrication uncertainties. 

\begin{figure}[tb]
    \centering
    \includegraphics[width=\linewidth]{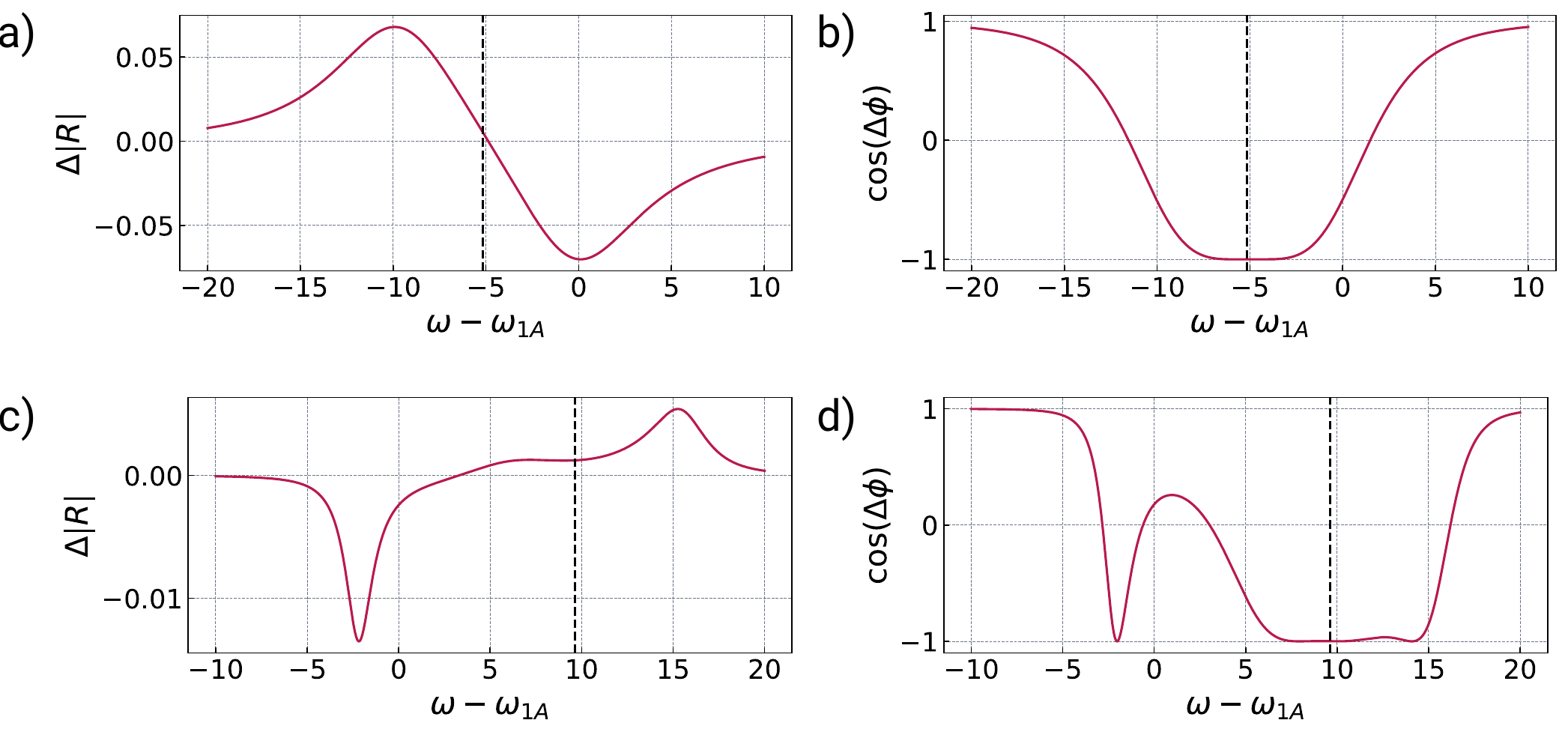}
    \caption{Visualization of the optimized amplitude and phase spectra for the SiV (a), (b) in the first row and SnV (c), (d), where the dashed black line denotes the optimized central frequency $\omega_0$. It is $\Delta\vert R\vert=\vert R_2\vert-\vert R_1\vert$ and $\Delta\phi=\phi_2-\phi_1$.}
    \label{fig:rob}
\end{figure}

In Fig. \ref{fig:rob} we visualize the optimized amplitude and phase spectra for the SiV and SnV, neglecting cross-talk. For the SnV we use optimized parameters at $\theta_{\rm dc}=\pi/2$ for illustrative purposes.
We observe flatness for the phase spectra for the SiV and SnV in a local environment of the optimized central frequency, which is due to robust optimization to broadband photons. We do not illustrate the spectra taking crosstalk into account because of the nonlinear nature of the upcoming terms in the Heisenberg-Langevin equations shown in \cite{strocka_memory_2025}. Nonlinear systems cannot be described by a single transfer function since a linear input-output relation can only describe LTI (linear time-invariant) systems \cite{Trentelman2002}. It is conceptually possible to illustrate such spectra for a damped input signal on a range of central frequencies. However, they depend on the photon-emitting duration, which is why we omit such visualizations here.

\subsection{Filtering}\label{app:filtering}
Assuming we use the quantum dot from \cite{schimpf_quantum_2021} and aim for narrower photon bandwidths, frequency filtering becomes essential. We model the filter as an empty, symmetric cavity with negligible losses. The transfer function mapping an input mode to a narrower output mode is the transmittivity of an empty cavity \cite{reiserer_cavity-based_2015}
\begin{align}
    F(\omega)=\frac{2\sqrt{\kappa_{f,l} \kappa_{f,r}}}{{\rm i}\omega+\kappa_f}
\end{align}
with the total cavity loss rate $\kappa_f=\kappa_{f,l}+\kappa_{f,r}+\kappa_{f,\rm loss}$, the loss rate on the left $\kappa_{f,l}$ and right hand side $\kappa_{f,r}$, respectively and the intrinsic losses $\kappa_{f,\rm loss}$. We assume negligible loss $\kappa_{f,\rm loss}=0$ and a symmetric cavity, i.e. $\kappa_{f,l}=\kappa_{f,r}=\kappa_f/2$. In that case the transmittivity reads
\begin{align}
    F(\omega)=\frac{\kappa_f}{{\rm i}\omega+\kappa_f}.
\end{align}
The incoming mode $S$ shown in Eq. \eqref{eq:spec} passes through the filter which leads to the output mode
\begin{align}\label{eq:fisp}
    S_f(\omega)=F(\omega)S(\omega)=\frac{\kappa_f}{{\rm i}\omega+\kappa_f} \frac{\epsilon_0}{{\rm i}\omega+\gamma/2}.
\end{align}
To get the bandwidth $\tilde{\gamma}$ after filtering we need the cavity bandwidth
\begin{align}
    \kappa_f=\frac{\tilde{\gamma}}{2}\sqrt{\frac{\gamma^2+\tilde{\gamma}^2}{\gamma^2-\tilde{\gamma}^2}}.
\end{align}
To model spin-photon interaction with crosstalk assuming the photon is emitted from a QD and passing subsequently through a Fabry-Perot interferometer we use the Heisenberg-Langevin equation shown in \cite{strocka_memory_2025} and the input mode in time domain. To derive the input mode after filtering in time domain $a_{\rm in}$ we evaluate the convolution between the functions
\begin{align}
    s(t)=\epsilon_0 e^{({\rm i}\omega_0-\gamma/2)t},\quad f(t)=e^{({\rm i}\omega_0-\kappa_f)t}.
\end{align}
It is
\begin{align}\label{eq:fisp_t}
    a_{\rm in}(t)=\int_0^t s(\tau)f(t-\tau)\,{\rm d}\tau=\frac{2\epsilon_0}{2\kappa_f-\gamma}f(t)\left(e^{(2\kappa_f-\gamma)/2 t}-1\right).
\end{align}
For optimizing a spin-photon CPHASE gate it is sufficient to optimize a cavity which is compatible with the QD from \cite{schimpf_quantum_2021}. To evaluate the fidelity and efficiency we assume that the incoming mode is given by Eq. \eqref{eq:fisp} when the model in frequency domain is sufficient and Eq. \eqref{eq:fisp_t} when time domain simulations are required to capture crosstalk effects and the G4V cavity as well as the emission frequency of the mode do not change. Since $2e_0/(2\kappa_f-\gamma)$ might be larger than $e_0$ one has to carefully choose $e_0$ to not produce unwanted driving between the states $\ket{1},\ket{A}$ and $\ket{2},\ket{B}$. Furthermore, the one has to multiply the factor $\mathcal{N}_{X,XX}$ from Eqs. \eqref{eq:IX1}-\eqref{eq:IX3} with $\kappa_f^2$. 

Assuming the QD from \cite{schimpf_quantum_2021} we apply filtering to make it compatible with the SiV at the magnetic field strength $B=0.3$ T and orientation $\theta_{\rm dc}=\pi/2$. We analyze that case because it leads to a spin splitting $\omega_s\approx 8$ GHz which is possible to address with microwaves to produce a spin $\pi/2$ rotation with state-of-the-art experimental techniques \cite{karapatzakis_microwave_2024}. At the field strength $B=1$ T it has not yet been experimentally realized and represents future work in microwave spin control for G4Vs. 

To optimize a spin-photon CPHASE gate for the SiV at $B=0.3$ T and $\theta_{\rm dc}=\pi/2$ we first optimize the SiV cavity and central frequency of the emitted photons. For that purpose we are restricted to assume a narrower bandwidth because spin-photon entanglement does not work at $\gamma_X=4.34$ GHz and $\gamma_{XX}=8.33$ GHz without filtering at the considered magnetic field strength. We assume a QD with $\gamma_X=\gamma_{XX}=1.56$ GHz without filtering to optimize a SiV cavity. By first optimizing the parameters in frequency domain as shown in App. \ref{app:no_cross} and subsequently improving the parameters when including crosstalk we find the fidelity $F=7.37\cdot 10^{-2}$ at the parameters $\kappa=17.82$ GHz, $\delta_c=\omega_{1A}-\omega_c=-38.56$ GHz, $\delta_0=\omega_{1A}-\omega_0=4.57$ GHz, $C_{1A},C_{2B}\approx 25$. We specifically restricted the cooperativity to $25$ because these are state-of-the-art SiV cavities \cite{bhaskar_experimental_2020} making the present simulations a meaningful contribution. When assuming the QD from \cite{schimpf_quantum_2021} with the rates $\gamma_X=4.34$ GHz and $\gamma_{XX}=8.33$ GHz with the above mentioned SiV cavity modeled by $(\kappa,\delta_c)$ and emission frequency governed by $\delta_0$ we find that $\tilde{\gamma}_X<4$ GHz and $\tilde{\gamma}_{XX}<7.3$ GHz is sufficient to achieve the infidelities $1-F_{\rm sp}<7.89\cdot 10^{-2}$ and efficiencies $\eta_{\rm sp}<0.3311$. That infidelity bound was used because it is on the same order of magnitude as the optimized spin-spin entanglement infidelity for the SiV at $B=1$ T and the broadband QD from \cite{schimpf_quantum_2021} with the rates $\gamma_X=4.34$ GHz and $\gamma_{XX}=8.33$ GHz which is shown in Tab. \ref{tab:summary}. Such a low infidelity is achieved even at so broad spectra using filtering subject to experimental constraints for the magnetic field strength $B=0.3$ T because of the steeper mode spectrum after filtering. That leads to compensation of fidelity loss even if the spectrum is such broad.


\cleardoublepage
\section{Expected Number of End-to-end Entanglement Establishments per Operational Cycle}
\label{App:expected}

Based on the definition, the expected number of end-to-end entanglement establishments over the whole repeater chain per operational cycle is given by
\begin{align}
    \mathrm{E}[n_{\rm end}]= & \sum_{l=1}^{m_{\rm s}} lP_{\rm end}(n_{\rm end}=l)
\end{align}
where $P_{\rm end}(n_{\rm end}=l)$ is the possibility to establish $l$ end-to-end entanglements in one cycle.
It can be rewritten as
\begin{align}
    \mathrm{E}[n_{\rm end}]= & \sum_{l=1}^{m_{\rm s}} \sum_{k=l}^{m_{\rm s}} P_{\rm end}(n_{\rm end}=l)\\
    =& \sum_{l=1}^{m_{\rm s}}  P_{\rm end}(n_{\rm end} \ge l)\;.
\end{align}

Assuming deterministic entanglement swapping, establishing at least $l$ end-to-end entanglements implies that each elementary link has also established at least $l$ entanglements. Consequently, we have
\begin{equation}
    P_{\rm end}(n_{\rm end} \ge l)= P_{\rm ele}^N(n_{\rm ele} \ge l)\;,
\end{equation}
where $P_{\rm ele}(n_{\rm ele} \ge l)$ is the possibility for an elementary link to establish at least $l$ elemental entanglements within one cycle. It is given by
\begin{align}
    P_{\rm ele}(n_{\rm ele} \ge l)= & \sum_{k=l}^{m_{\rm s}} P_{\rm ele}(n_{\rm ele} = l) \\
    = & \sum_{k=l}^{m_{\rm s}} \frac{m_{\rm s}!}{(m_{\rm s}-k)!k!} p_{\rm arm}^k(1-p_{\rm arm})^{m_{\rm s}-k} \;.
\end{align}

\section{Optimized parameters}
\label{app:opt_m_N}

The optimized values of $\{N, n_{\rm loa}, n_{\rm dis,n}, n_{\rm dis,e}\}$ and the corresponding secret-key rate $R_{sk, opt}$ for different $\{F_{\rm sp}, L, m, \varepsilon_{\rm n-n}\}$ values are presented in Tab.~\ref{tab:opt_para1}. Note that the optimum $N$, $n_{\rm loa}$, $n_{\rm dis,n}$, and $n_{\rm dis,e}$ are searched within a wide enough range and the smallest possible step of one. However, the the searching range of $n_{\rm loa}$ is $\{\mathrm{round}\{\exp[\ln(20)+n\tau]\}\}$, where $\mathrm{round}(\cdot)$ denotes rounding to the nearest integer, $n=0, 1, 2, \cdots, 99$, and $\tau=[\ln(20{,}000)-\ln(20)]/99$.

\begin{table}[htbp]
\centering
\caption{Optimized values ($N$, $n_{\rm loa}$, $n_{\rm dis,n}$, $n_{\rm dis,e}$) and the corresponding secret-key rates ($R_{sk, opt}$).}
\label{tab:opt_para1}
\begin{tabular}{c c c | c c c c c | c c c c c}
\hline
\multirow{2}{*}{$\varepsilon_{\rm n-n}$} & \multirow{2}{*}{$F_{\rm sp}$} & \multirow{2}{*}{$L$} & 
\multicolumn{5}{c|}{$m=100$} & 
\multicolumn{5}{c}{$m=1000$} \\
 & & & $N$ & $n_{\rm loa}$ & $n_{\rm dis,n}$ & $n_{\rm dis,e}$ & $R_{\rm sk}$ & 
$N$ & $n_{\rm loa}$ & $n_{\rm dis,n}$ & $n_{\rm dis,e}$ & $R_{\rm sk}$ \\
\hline
0.01 & 0.95 & 100  & 2 & 20  & 0 & 0 & $2.14\times 10^{4}$ & 2 & 20  & 0 & 0 & $2.31\times 10^{5}$ \\
0.01 & 0.95 & 200  & 2 & 40  & 0 & 0 & $3.64\times 10^{3}$ & 2 & 40  & 0 & 0 & $4.20\times 10^{4}$ \\
0.01 & 0.95 & 300  & 2 & 114 & 0 & 0 & $6.33\times 10^{2}$ & 2 & 107 & 0 & 0 & $8.35\times 10^{3}$ \\
0.01 & 0.95 & 400  & 4 & 114 & 0 & 1 & $1.66\times 10^{2}$ & 4 & 114 & 0 & 1 & $2.29\times 10^{3}$ \\
0.01 & 0.95 & 500  & 4 & 200 & 0 & 1 & 57.4                & 4 & 174 & 0 & 1 & 927 \\
0.01 & 0.95 & 600  & 5 & 214 & 0 & 2 & 19.6                & 5 & 174 & 0 & 2 & 416 \\
0.01 & 0.95 & 700  & 6 & 214 & 0 & 2 & 6.54                & 5 & 247 & 0 & 2 & 196 \\
0.01 & 0.95 & 800  & 6 & 283 & 0 & 2 & 2.34                & 5 & 350 & 0 & 2 & 89.6 \\
0.01 & 0.95 & 900  & 6 & 462 & 0 & 2 & 0.692               & 6 & 304 & 0 & 2 & 46.0 \\
0.01 & 0.95 & 1000 & 6 & 611 & 0 & 2 & 0.157               & 6 & 431 & 0 & 2 & 22.9 \\
0.01 & 0.95 & 1100 & 7 & 495 & 0 & 3 & 0.0303              & 6 & 611 & 0 & 2 & 10.9 \\
0.01 & 0.95 & 1200 & 7 & 753 & 0 & 3 & 0.00614             & 6 & 866 & 0 & 2 & 4.79 \\
\hline
0.01 & 0.98 & 100  & 2 & 20  & 0 & 0 & $4.62\times 10^{4}$ & 2 & 20  & 0 & 0 & $5.00\times 10^{5}$ \\
0.01 & 0.98 & 200  & 2 & 40  & 0 & 0 & $7.88\times 10^{3}$ & 2 & 40  & 0 & 0 & $9.09\times 10^{4}$ \\
0.01 & 0.98 & 300  & 2 & 114 & 0 & 0 & $1.37\times 10^{3}$ & 7 & 46  & 0 & 1 & $1.88\times 10^{4}$ \\
0.01 & 0.98 & 400  & 7 & 70  & 0 & 1 & 562                 & 7 & 65  & 0 & 1 & $9.92\times 10^{3}$ \\
0.01 & 0.98 & 500  & 8 & 87  & 0 & 1 & 278                 & 8 & 75  & 0 & 1 & $5.59\times 10^{3}$ \\
0.01 & 0.98 & 600  & 8 & 107 & 0 & 1 & 138                 & 8 & 100 & 0 & 1 & $3.31\times 10^{3}$ \\
0.01 & 0.98 & 700  & 9 & 114 & 0 & 1 & 66.6                & 8 & 132 & 0 & 1 & $1.99\times 10^{3}$ \\
0.01 & 0.98 & 800  & 11& 123 & 0 & 2 & 33.7                & 8 & 162 & 0 & 1 & $1.19\times 10^{3}$ \\
0.01 & 0.98 & 900  & 11& 151 & 0 & 2 & 17.7                & 9 & 162 & 0 & 1 & 751 \\
0.01 & 0.98 & 1000 & 12& 141 & 0 & 2 & 8.94                & 9 & 214 & 0 & 1 & 472 \\
0.01 & 0.98 & 1100 & 12& 174 & 0 & 2 & 4.38                & 11& 174 & 0 & 2 & 322 \\
0.01 & 0.98 & 1200 & 12& 230 & 0 & 2 & 1.86                & 11& 200 & 0 & 2 & 219 \\
\hline
0.10 & 0.95 & 100  & 2 & 20  & 0 & 0 & 500                 & 2 & 20  & 0 & 0 & 5410 \\
0.10 & 0.95 & 200  & 2 & 40  & 0 & 0 & 85.2                & 2 & 40  & 0 & 0 & 983 \\
0.10 & 0.95 & 300  & 2 & 114 & 0 & 0 & 14.8                & 2 & 107 & 0 & 0 & 195 \\
0.10 & 0.95 & 400  & 2 & 283 & 0 & 0 & 2.11                & 2 & 264 & 0 & 0 & 37.4 \\
0.10 & 0.95 & 500  & 2 & 866 & 0 & 0 & 0.197               & 2 & 655 & 0 & 0 & 6.08 \\
0.10 & 0.95 & 600  & 2 & 2000& 0 & 0 & 0.0122              & 2 & 1739& 0 & 0 & 0.695 \\
0.10 & 0.95 & 700  & 2 & 5312& 0 & 0 & 0.000570            & 2 & 4954& 0 & 0 & 0.0466 \\
0.10 & 0.95 & 800  & 2 &14110& 0 & 0 & $2.18\times 10^{-5}$& 2 &14110& 0 & 0 & 0.00207 \\
0.10 & 0.95 & 900  & 2 &20000& 0 & 0 & $5.98\times 10^{-7}$& 2 &20000& 0 & 0 & $6.10\times 10^{-5}$ \\
0.10 & 0.95 & 1000 & 2 &20000& 0 & 0 & $9.06\times 10^{-9}$& 2 &20000& 0 & 0 & $9.15\times 10^{-7}$ \\
0.10 & 0.95 & 1100 & 2 &18652& 0 & 0 & $1.06\times 10^{-10}$& 2&20000& 0 & 0 & $1.03\times 10^{-8}$ \\
0.10 & 0.95 & 1200 & 2 &16223& 0 & 0 & $1.31\times 10^{-12}$& 2&20000& 0 & 0 & $1.09\times 10^{-10}$ \\
\hline
0.10 & 0.98 & 100  & 2 & 20  & 0 & 0 & $1.81\times 10^{4}$ & 2 & 20  & 0 & 0 & $1.96\times 10^{5}$ \\
0.10 & 0.98 & 200  & 2 & 40  & 0 & 0 & $3.09\times 10^{3}$ & 2 & 40  & 0 & 0 & $3.57\times 10^{4}$ \\
0.10 & 0.98 & 300  & 2 & 114 & 0 & 0 & 538                 & 2 & 107 & 0 & 0 & 7091 \\
0.10 & 0.98 & 400  & 2 & 283 & 0 & 0 & 76.6                & 2 & 264 & 0 & 0 & 1355 \\
0.10 & 0.98 & 500  & 2 & 866 & 0 & 0 & 7.16                & 2 & 655 & 0 & 0 & 221 \\
0.10 & 0.98 & 600  & 2 &2000 & 0 & 0 & 0.444               & 2 &1739 & 0 & 0 & 25.2 \\
0.10 & 0.98 & 700  & 2 &5312 & 0 & 0 & 0.0207              & 2 &4954 & 0 & 0 & 1.69 \\
0.10 & 0.98 & 800  & 2 &14110& 0 & 0 & 0.000791            & 2 &14110& 0 & 0 & 0.0751 \\
0.10 & 0.98 & 900  & 2 &20000& 0 & 0 & $2.17\times 10^{-5}$& 2 &20000& 0 & 0 & 0.00221 \\
0.10 & 0.98 & 1000 & 2 &20000& 0 & 0 & $3.29\times 10^{-7}$& 2 &20000& 0 & 0 & $3.32\times 10^{-5}$ \\
0.10 & 0.98 & 1100 & 2 &18652& 0 & 0 & $3.84\times 10^{-9}$& 2 &20000& 0 & 0 & $3.73\times 10^{-7}$ \\
0.10 & 0.98 & 1200 & 2 &16223& 0 & 0 & $4.75\times 10^{-11}$& 2&20000& 0 & 0 & $3.97\times 10^{-9}$ \\
\hline
\end{tabular}
\end{table}

\end{appendix}
\end{document}